\documentclass[11pt,a4paper]{article}

\usepackage{amsmath,amssymb}
\usepackage{epsfig,graphicx}
\usepackage{subfigure}
\usepackage{graphicx}
\usepackage{rotating}
\usepackage{cancel}
\usepackage{bm}
\usepackage{color}
\usepackage{comment}
\usepackage{cite}
\usepackage{psfrag}
\usepackage{hyperref}

\renewcommand\({\left(}
\renewcommand\){\right)}
\renewcommand\[{\left[}

\newcommand{\Slash}[1]{{\ooalign{\hfil#1\hfil\crcr\raise.167ex\hbox{/}}}}

\newcommand{\beq}{\begin{equation}}  \newcommand{\eeq}{\end{equation}}
\newcommand{\bef}{\begin{figure}}  \newcommand{\eef}{\end{figure}}
\newcommand{\bec}{\begin{center}}  \newcommand{\eec}{\end{center}}
  
\newcommand{\laq}[1]{\label{eq:#1}}  

\newcommand{\Eq}[1]{Eq.~(\ref{eq:#1})}

\newcommand{\eq}[1]{(\ref{eq:#1})}

\newcommand{\vev}[1]{ \left\langle {#1} \right\rangle }

\def\({\left(}
\def\){\right)}

\def\O{\mathcal{O}}

\newcommand{\OR}{~{\rm or}~}
\newcommand{\AND}{~{\rm and}~}
\newcommand{\EV}{ {\rm \, eV} }
\newcommand{\KEV}{ {\rm \, keV} }
\newcommand{\MEV}{ {\rm \, MeV} }
\newcommand{\GEV}{ {\rm \, GeV} }
\newcommand{\TEV}{ {\rm \, TeV} }

\def\a{\alpha}

\def\d{\delta}
\def\e{\epsilon}
\def\f{\phi}
\def\g{\gamma}

\def\l{\lambda}

\def\n{\nu}

\def\s{\sigma}

\def\D{\Delta}
\def\G{\Gamma}

\def\L{\Lambda}
\def\F{\Phi}

\def\tl{\tilde}
\def\*{\dagger}

\newcommand{\exclude}[1]{}

\def\beq{\begin{equation}}
\def\eeq{\end{equation}}

\begin{document}
\numberwithin{equation}{section}
\title{\vspace*{1.5cm}Boosted Neutrinos and Relativistic Dark Particles as Messengers from Reheating
\vspace{0.8cm} 
\Large{\textbf{
}}}

\author{Joerg Jaeckel$^{1}$ and Wen Yin$^{2,3}$\\[2ex]
\small{\em $^1$Institut f\"ur theoretische Physik, Universit\"at Heidelberg,} \\
\small{\em Philosophenweg 16, 69120 Heidelberg, Germany}\\[0.5ex]  
\small{\em $^2$Department of Physics, Faculty of Science, The University of Tokyo, 
} \\
\small{\em Bunkyo-ku, Tokyo 113-0033, Japan}\\[0.5ex]  
\small{\em $^3$Department of Physics, KAIST,  Daejeon 34141, Korea}\\[0.8ex]}

\date{}
\maketitle

\begin{abstract}
Usually information from early eras such as reheating is hard to come by. In this paper we argue that, given the right circumstances, right-handed sterile neutrinos decaying to left-handed active ones at relatively late times can carry information from reheating by propagating freely over the thermal history. For not too small mixing angles, suitable right-handed neutrino masses are around $\O(\MEV-\GEV)$. 
We identify the typical spectra and argue that they provide information on the ratio of the inflaton mass to the reheating temperature. 
This primordial neutrino signal can be strong enough that it can be detected in IceCube. More speculatively, for a reheating temperature and inflaton mass satisfying 
$
T_R=\O(1-100)\MEV, \AND m_\f=\O(10^{16-19})\GEV$ they may even explain the observed PeV events. 
Also more general relativistic dark particles can play the role of such messengers,
potentially not only allowing for the PeV events but also alleviating the $H_0$-tension. 
\noindent
\end{abstract}

\newpage

\section{Introduction}
Explaining the flatness and homogeneity, but also providing the small inhomogeneities that seed structure formation, inflation~\cite{Starobinsky:1980te,Guth:1980zm,Sato:1980yn,Linde:1981mu,Albrecht:1982wi} is widely accepted as a central phase in the evolution of the Universe.
Inflation ends with the decay of the inflaton, reheating the Universe and thereby starting the thermal history. Cosmological observations, e.g. by the Planck satellite~\cite{Akrami:2018odb}, agree well with this scenario.
However, up to now, there is still very little direct experimental or observational evidence of details of the reheating era.
Since after the reheating the Universe is hot and dense, any information is lost by the 2nd law of thermodynamics unless there is a weakly-coupled messenger. A good messenger should freely travel over the thermal history and reach Earth at the speed of light. Then information about reheating can be experimentally accessed through its spectrum.

An important example of a messenger that survives from inflation are gravitational waves. They could provide information on the pre-heating phase after inflation which can feature parametric resonances, bubble formation or objects such as oscillons giving rise to gravitational waves, cf., e.g.~\cite{Tashiro:2003qp,Easther:2006gt,GarciaBellido:2007af,Dufaux:2007pt, Huang:2011gf,Hebecker:2016vbl,Kitajima:2018zco,Amin:2018kkg,Lozanov:2019ylm,Sang:2019ndv}. 
The frequency and spectra of the gravitational waves depend on the shape of the inflaton potential and inflaton couplings, the graviton can thus be a messenger of the (p)reheating. 

Another important option is that reheating from the decays of moduli in large volume models of the axiverse can provide a large amount of dark radiation composed of relativistic axion-like particles~\cite{Cicoli:2012aq,Higaki:2012ar,Conlon:2013isa,Hebecker:2014gka}. 
Through the gravitational interaction, the cosmic microwave background (CMB) and the baryonic acoustic oscillations can be affected, allowing for a test of this possibility. Moreover, the energy spectrum of these axion-like particles is linked to the masses of the moduli responsible~\cite{Conlon:2013isa}. If the coupling to photons is sufficiently strong the axion-like particles may even be visible in IAXO~\cite{Armengaud:2019uso}. 
Its measurement could therefore be a direct probe of reheating~\cite{Conlon:2013isa,Armengaud:2019uso}.\footnote{Reheating may also be probed in the laboratory by searching for light inflatons in accelerator based experiments. The decay in such an experiment is through the same process relevant for reheating and would in this way be a test of this period in the evolution of the Universe~\cite{Bezrukov:2009yw,Takahashi:2019qmh,Okada:2019opp}. That said, this only tests for the existence of a suitable particle and does not directly show that this process was realized during the evolution of the Universe.}

In spirit our approach is the same as the scenario where moduli decay into axion-like particles~\cite{Cicoli:2012aq,Higaki:2012ar,Conlon:2013isa,Hebecker:2014gka,Armengaud:2019uso}.
However, instead of axion-like particles we study the possibility that neutrinos or other relativistic and sufficiently dark particles act as suitable messengers. 
We show that these particles, produced 
from inflaton (or modulus) decays, can propagate to the present Universe without being scattered by the ambient thermal plasma.
Thereby they can carry the information about reheating until today,
and can be tested from ground-based experiments. 
In particular we discuss that ``primordial" neutrinos from reheating can reach Earth in a simple renormalizable model with a right-handed neutrino in the mass range of $\O(\MEV-\GEV).$
We identify the typical spectra and discuss the testability in experiments for cosmic-ray neutrinos. The spectrum then can give information on the mass of the inflaton in relation to the temperature at the time of its decay.
 
Such a scenario could be within reach. For example, when the inflaton mass, $m_\f$ and reheating temperature, $T_R$, satisfy $m_\f=\O(10^{16-19})\GEV,$ and $T_R=\O(1\MEV-1\GEV),$ the primordial neutrino may even explain the PeV events in IceCube. 
A boosted sterile neutrino can also be a messenger of reheating and explains the PeV events through its decay to energetic neutrinos. 
Since the relativistic component can contribute to the dark radiation with $\Delta N_{\rm eff}=\O(0.1)$, the $H_0$-tension can be alleviated simultaneously.

A general stable messenger is also discussed, including the relevant condition for it to travel freely until today, as well as the experimental event rates it could generate. Indeed it may even be possible that the same particle that acts as a stable messenger may, produced differently, simultaneously plays the role of the dominant 
dark matter (DM).

Cosmic-rays of neutrinos and boosted DM from heavy long-lived particle decays in late epochs have been discussed in various contexts~\cite{Feldstein:2013kka,Esmaili:2013gha,Ema:2013nda, Higaki:2014dwa,Rott:2014kfa, Ema:2014ufa,Dudas:2014bca,Murase:2015gea,Dev:2016qbd,Hiroshima:2017hmy,Bhattacharya:2014yha,Kopp:2015bfa,Cui:2017ytb}. Most of them consider the production of high energy neutrinos and weakly coupled particles from the decay of DM today. However, the authors of Refs.~\cite{Ema:2013nda, Ema:2014ufa} also consider 
a heavy particle decaying to neutrinos at a time later than a redshift of $10^6$. For larger redshift, the Universe is opaque to the neutrinos carrying energy greater than PeV scales today (see the discussion in~\cite{Ema:2014ufa} where this case has been investigated).
In this paper we mostly focus on decays that happen before a red-shift of $10^6$. In particular we view the neutrinos or some highly boosted DM particle as a messenger of the ``reheating phase" originating from the decays of a heavy, non-relativistic inflaton decaying before big-bang nucleosynthesis (BBN).  
The main phenomenological difference from the previous works are the spectra and the angular dependence.
In particular, cosmic-rays from non-relativistic DM decaying today tend to come from the galactic center or other concentrations of DM. In contrast, in our case the decays happen very early (before BBN). Therefore, they are almost isotropic with $\O(10^{-5})$ anisotropies linked to those in the CMB. 

To have a concrete scenario we consider the inflaton to decay into right-handed neutrinos which, at a later time, decay to active neutrinos. 
This allows the energetic active neutrinos to reach Earth without being scattered and to have a theoretical consistent inflaton coupling.  
The subsequent decays of the boosted right-handed neutrinos to active ones then provide typical spectra, which are different from those obtained in the previous works mentioned above for other situations.

Let us briefly outline our plan for the paper. In the next section we discuss a model for reheating, and determine the parameter region where primordial neutrinos from reheating can reach Earth. In Sec.\,\ref{sec:3} we obtain the primordial neutrino flux and investigate its phenomenology. In Sec.\,\ref{sec:H0}, the flux and phenomenology of more general relativistic dark particles originating from inflaton decays are discussed. The last section is devoted to a short discussion and conclusions.

\section{Primordial neutrinos from inflaton decays}
 If a decay product of the inflaton, $\f$, which we take as a real scalar field, is sufficiently weakly coupled to the thermal bath of standard model (SM) particles, its energy distribution and therefore its spectrum may be preserved until today. The spectrum can then carry information about the reheating phase.
 As a concrete example, we study the situation that energetic (right-handed) neutrinos from inflaton decays reach Earth. We focus on high energy neutrinos for two main reasons. First, there exists a large volume detector, 
IceCube~\cite{icecube}, suitable for their detection. Second, the background flux of very highly energetic cosmic-ray neutrinos is less than that in the lower energy range.

\subsection{Reheating and production of neutrinos}

Let us first discuss reheating from the decay of a heavy inflaton, $\phi$, with mass $m_\phi$. In Appendix~\ref{ap1}, we review models for successful inflation and investigate conditions for the absence of tunings introduced by the 
interactions given below in \eq{Lag}.

Introducing a right-handed neutrino, $N$, we can generate a mass term for the neutrino e.g. via the see-saw mechanism~\cite{Yanagida:1979as,Glashow:1979nm, GellMann:1980vs,Minkowski:1977sc,Mohapatra:1979ia}. This provides a suitable particle for our inflaton to decay to. 
Then the most general renormalizable Lagrangian with leptons is 
\begin{equation}
\laq{Lag}
{\cal L}=-\frac{1}{2}(M_N+\l\d\f) \bar{N}^c N- y h \bar{N} \hat{P}_L L- A \d \f |h|^2-\frac{\l_p}{2} \d\f^2 |h|^2-\frac{m_\f^2}{2} \d\f^2+\cdots
\end{equation}
where $L$ a left-handed lepton doublet, and $h$ the Higgs doublet. 
$( \l, y, A, \l_p)$ are couplings. In particular they include the neutrino Yukawa coupling $y$, $M_N$ is the right-handed neutrino mass, and $\cdots$ indicate self-interaction terms for $\d\f$, and the interactions amongst SM particles.
We have defined $\d\f\equiv \f-\vev{\f}$ so that $\d\f=0$ is the minimum of the potential where $\vev{\f}$ is the vacuum expectation value of $\f$. 
Here and hereafter, for simplicity, we consider only a single flavor of leptons, unless otherwise stated. Our analysis can be easily extended to the multi-flavor case.

After inflation, $\f$ starts to oscillate around the potential minimum. The oscillation energy dominates the energy density of the Universe. During the oscillation, the inflaton decays dominantly to
\begin{align}
\laq{fNN}\f\to N N, \\ 
\f\to hh,
\end{align}
with the decay rates\footnote{In the decay rate to the Higgs bosons we include the contributions of the Goldstone bosons. This approximately takes into account decays into longitudinal gauge bosons.},
\begin{align}
\laq{decayrate0}
\Gamma_{\f \to hh} &\simeq \frac{A^2}{\pi m_\phi},\\~~~\Gamma_{\f \to NN} &\simeq \frac{\lambda^2 }{4\pi}m_\f.
\end{align}
In both cases, for simplicity, we have taken the massless limit of the decay products. 
The total decay is the sum,
\begin{equation}
\Gamma_\f\approx \G_{\f\to NN}+\G_{\f\to hh}.
\end{equation}
It reheats the Universe\footnote{We neglect the effect from parametric/tachyonic growth since the order parameter for a broad resonance, $q\equiv A^2\rho_\f/m_\phi^6$, is highly suppressed in our parameter space under consideration (see, e.g., Ref.~\cite{Amin:2018kkg} for the case of parametric/tachyonic resonance for the $\phi |h|^2$ coupling).} to a temperature of the order of, 
\beq 
\laq{reht} 
T_R \equiv \(\frac{g_\star\pi^2}{90}\)^{-1/4}\sqrt{\G_\f M_{\rm pl}},
\eeq 
where $g_\star$ is the number of relativistic degrees of freedom (for which we use the values from Ref.~\cite{Husdal:2016haj}). 
For instance, if the dominant decay is via the coupling $A$, i.e. $\G_{\f\to NN} \ll \G_{\f \to hh} \leftrightarrow \l \ll A/m_\f,$
we obtain\footnote{With $m_\f \gg T_R$, a baryon asymmetry production with higher dimensional terms, such as an neutron-antineutron oscillation term, can be enhanced~\cite{Asaka:2019ocw}.} 
\begin{equation}
T_R \sim  300\MEV \(\frac{A}{1\GEV}\)\(\frac{M_{\rm pl}}{m_\f}\)^{1/2}.
\end{equation} 
Even if the inflaton is as heavy as $m_\phi \sim M_{\rm pl},$ the reheating temperature can be low for small $A$. This hierarchy between $T_R$ and $m_\phi$ can be accommodated in certain models (see Appendix~\ref{ap1}).

\subsection{Propagating neutrinos from \texorpdfstring{$\f$}{phi} decays} 
Let us discuss the conditions under which a boosted neutrino from inflaton decays travels freely to Earth. 
The neutrinos carrying the information from reheating originate from a two step process,
\beq
\f \to NN\to \nu\nu+\cdots. 
\eeq
It is known that SM particles with an energy around $\sim m_\f$ can be thermalized with $m_\f \gg T_R\gtrsim 1\MEV$ through a ``bottom-up" thermalization~\cite{Allahverdi:2002pu, Kurkela:2011ti, Harigaya:2013vwa} (see also Refs.\cite{Davidson:2000er,Baier:2000sb}), where the slowest splitting process has a rate $ \sim 10^{-4}\sqrt{T_R^3/m_\f},$ which is much faster than the expansion of the Universe for $ T_R\ll m_\f \lesssim M_{\rm pl}$.  Therefore $h$ from $\f\to hh$ are soon thermalized. 
In the remainder of the paper, we assume 
\beq
\G_{\f\to NN}\lesssim \G_{\f\to hh},
\eeq
and use that the SM particles are thermalized promptly. 

\paragraph{Survival of the signal I: No scattering of $N$}
Since the SM particles can be thermalized promptly around the last stage of the reheating, the Universe features an ambient plasma of SM particles when $H\lesssim \G_\f.$
In the case $\f$ decays into an $N$ pair,  $N$ should rarely scatter with the ambient plasma in order to propagate. It should also not decay too soon after the reheating otherwise the products of the SM particles would be thermalized. 
In the phase with broken electroweak symmetry, i.e. when the temperature of the thermal bath is $T\ll 100\GEV$, 
$N$ mixes with the active neutrino $\nu$ with a mixing angle
\begin{equation}
\theta \equiv \frac{y v}{M_N},
\end{equation} 
where $v\approx 174\GEV$ is the Higgs vacuum expectation value. 
$N$ interacts with the plasma less frequently than a neutrino does since $|\theta|$ will turn out to be $\ll 1$. 

\begin{figure}[t!]
\begin{center}  
\includegraphics[width=105mm]{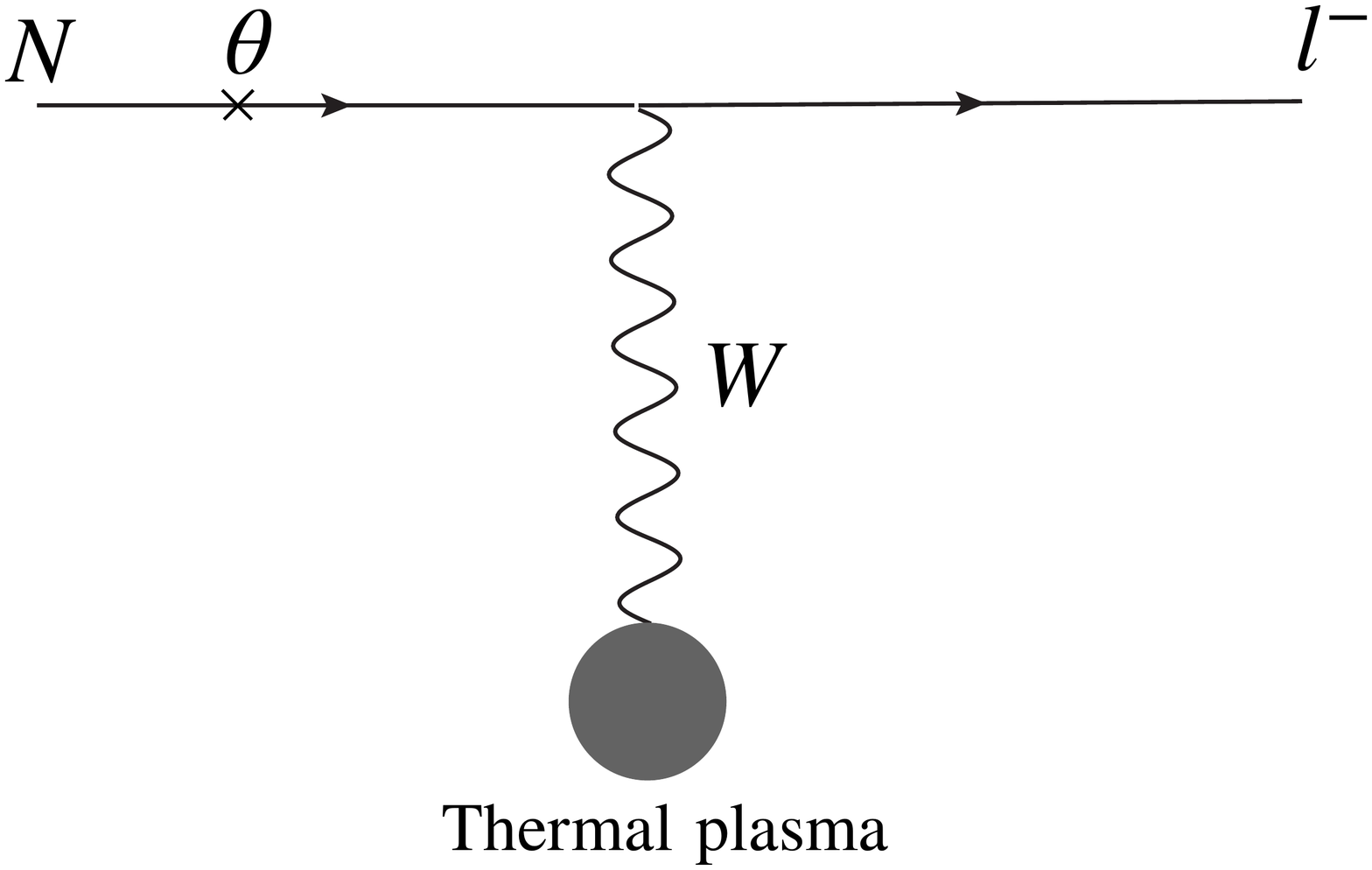}
\end{center}
\caption{Relevant diagram for the thermalization of $N$. 
}
\label{fig:diag1}
\end{figure}

\begin{figure}[t!]
\begin{center}  
\includegraphics[width=145mm]{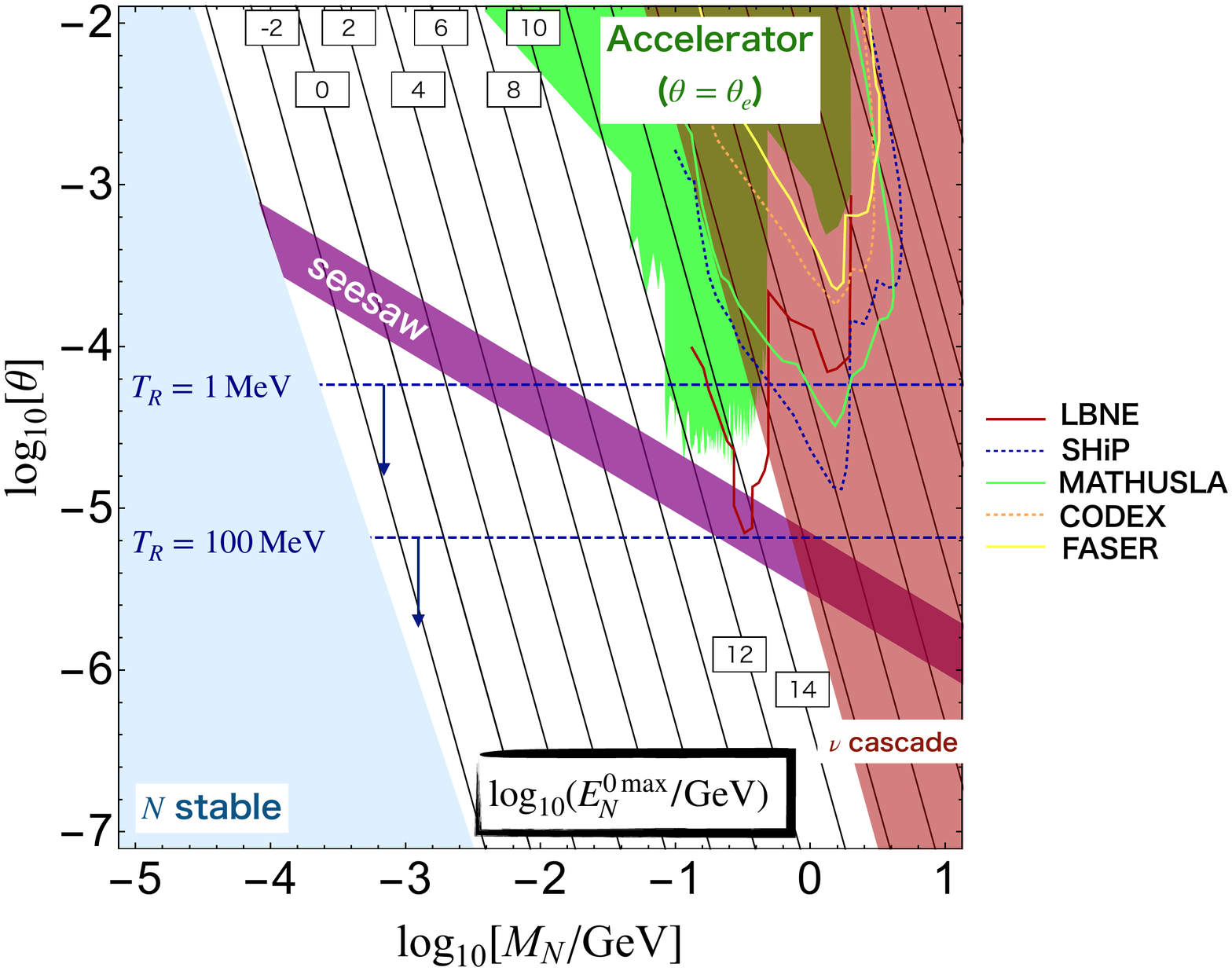}
\end{center}
\caption{Parameter region where primordial neutrinos produced in inflaton decays reach Earth without being scattered by the thermal plasma. 
We fix $C_{\rm scat}=1$, $C_{\rm decay}=3$.
The black lines indicate the energy, $E^0_N$, that the right-handed neutrino would have today. (We note that energies above $E^0_N\sim 10^{8}\,{\rm GeV}$ require, $\f$ to have a mass above the Planck scale, cf.~\eq{E1}).
The region is restricted by the conditions of free-propagation of $N$ [below the blue dashed lines for $T_R=1\MEV$ and $100\MEV$], $N$ decays until today [$E^0_N$ smaller than the contour value, right to the blue shaded region], and absence of cascade of $\nu$ from the decay of $N$ [left to the red shaded region]. 
Also shown is the ``see-saw relation" by the purple band corresponding to $\theta^2 M_N= [0.009-0.05]\EV.$  
The accelerator bounds obtained in~\cite{Beacham:2019nyx,deGouvea:2015euy,NA62:2020mcv}, using data from PS191~\cite{Bernardi:1987ek}, CHARM~\cite{Bergsma:1985is}, TRIUMF~\cite{Britton:1992pg,Britton:1992xv}, and NA62~\cite{NA62:2020mcv} are shown as the green-shaded region. The future reach~\cite{Beacham:2019nyx,Adams:2013qkq} of accelerator experiments are LBNE~\cite{Adams:2013qkq} [red solid line], MATSUSLA200~\cite{Chou:2016lxi} [grean solid line], FASER2~\cite{Kling:2018wct} [yellow solid line], CODEX-b~\cite{Gligorov:2017nwh} [orange dotted line] and SHiP~\cite{Alekhin:2015byh} [blue dotted line].  }
\label{fig:para}
\end{figure}

A typical interaction rate of $N$ can be calculated, 
for instance, from a t-channel scattering mediated by a W-boson as in Fig.\,\ref{fig:diag1}. 
The cross section can be approximated by  
\begin{equation}
\laq{crossN}
\s_{{\rm scat},N}^{W}\sim \theta^2 \times 4\pi \a_{2}^2 \int_{m^2_{W,Z}}{\frac{dt}{t^2} }\sim G_F \theta^2 g_2^2.
\end{equation}
Here we cut off the integral by using the weak boson mass. 
This cross section is suppressed by $\theta^2$ compared to the SM neutrino's cross section.
The scattering rate is obtained as
\begin{equation}
\G_N^{\rm scat} \sim \s_{{\rm scat},N}^{W}\times n_{\rm th}.
\end{equation}
This places a clear constraint on the mixing because if this happens soon after reheating the produced SM leptons would soon be thermalized and loose all information
(cf. Fig.~\ref{fig:diag1}).
The condition that ensures that a sufficient number of $N$ survive is given by $\G^N_{\rm scat}\lesssim 4H\equiv 4\sqrt{g_\star(T) \pi^2/90} \times T^2/M_{\rm pl}$.\footnote{Here we have compared the last two terms of the Boltzmann equation $d\rho_N/dt +4 H\rho_N+\G_N^{\rm scat} \rho_N=0$, where $\rho_N$ is the energy density of $N$. By solving the Boltzmann equations including the ones for $\rho_\f$ and the SM radiation, we have checked that $\rho_N\approx (5-10)\% \times B_{\f \to NN}\times 3 M_{\rm pl}^2 H^2 $ remains when $4H=\G_N^{\rm scat}$ is set at $T=T_R$. The fraction of $5-10\%$ depends on $T_R\approx\{1\MEV-1\GEV\}$. }
If this is fulfilled, $N$ propagates freely until its decay. 
As a result, the Universe is transparent for the energetic $N$ if 
\beq
\laq{th2}
\theta \lesssim 3\times 10^{-5} \(\frac{g_\star{(T_R)}}{11}\)^{1/4} \sqrt{\frac{5\MEV}{T_R}},
\eeq
where we take $T=T_R$ in the previous inequality since $\G^N_{\rm scat}$ decreases faster than $H$ due to the expansion.
There is also resonant production of $Z,W$ when the energy $E_N$ satisfies $E_N\sim m_{Z,W}^2/T$. The cross section for this is of the order of \eq{crossN}, and so the inclusion of this effect does not qualitatively change our conclusion. When $T_R\gg 100\GEV$ in the symmetric phase, $N$ can loose its energy through the Higgs exchange process of $N+ t\to t+ \nu $ etc. 
The corresponding scattering rate of $N$ with the thermal plasma can be calculated to be (c.f. \cite{Akhmedov:1998qx})
\beq
\G_N^{\rm scat, EW}\sim \frac{9 y_t^2y^2}{64\pi^3} \frac{T^2}{E_N}.
\eeq
This decreases as $\propto a^{-1}$ which is slower than the decrease of the Hubble parameter. 
By comparing $\G_N^{\rm scat,EW}$ with $4H$ at $T= 100\GEV$ we find that the sterile neutrino can propagate freely until $T\sim 100\GEV$ from any $T_R\gg 100\GEV$ fulfilling,  
\beq
\laq{ycons}
y\lesssim 0.3\sqrt{\frac{E_N}{10^{14}\GEV}},
\eeq
where $E_N$ is the energy of $N$ at $T\sim 100\GEV.$ In particular, if $\theta\lesssim 10^{-8}$ 
to satisfy \eq{th2} at $T \lesssim 10^2\GEV$ and if $M_N$ is small enough,  \Eq{ycons} can be easily satisfied.
Therefore, in principle, the neutrino can be a messenger with $T_R\gg 100\GEV$. 
However, as we will see, the messenger would have too low energy today from \Eq{E1}, given that $m_\f<10^{19}\GEV$, and $N$ cannot satisfy the see-saw relation (cf. Fig.\,\ref{fig:para}). Therefore, we will mainly discuss $T_R\ll 100\GEV$  and $N$ fulfilling condition \eq{th2} in this section.

\paragraph{$N$ decay}
If the Universe is transparent to it, $N$ remains energetic until late times.  
However, we also have to consider possible decays of $N$. 
The decay rate in cosmic time i.e. including the relativistic $\gamma$-factor $M_{N}/E_{N}$, is
\begin{align}\laq{decayrate}
\G^{\rm decay}_N\approx \frac{M_N}{E_N} \times\left\{\begin{array}{ll}{\displaystyle{ \frac{y^2 M_N}{8\pi}}~ (M_N\gg 100\GEV)}\\ 
&\\ 
\displaystyle{
C_{\rm decay}\frac{G_F^2 M_N^5}{192\pi^3}\theta^2~ (M_N\ll 100\GEV)} 
\end{array} \right..
\end{align}
In the first row, we give the decay rate to a Higgs doublet and a lepton doublet, and in the second the rate into leptons or hadrons via mixing. 
$C_{\rm decay}[M_N]=\O(1-10)$ parametrizes the sum of the various available hadronic channels depending on the mass scale of $M_N$~\cite{Gorbunov:2007ak,Dib:2018iyr}. 
In particular, when $1\MEV \lesssim M_N\lesssim 140\MEV$ the hadronic decays are forbidden and we obtain $C_{\rm decay}\approx 2-2.3$ depending on the flavor that $N$ mixes with. 
For $ M_N\lesssim 1\MEV$, the decays to electrons are forbidden and $C_{\rm decay}\approx 2$ independent of the flavor with which it mixes~\cite{Ruchayskiy:2012si}.
When the decay of $N$ happens at the radiation dominant epoch, one obtains the production temperature of the active neutrino from the condition $\G^{\rm decay}_N\sim 4H$,\footnote{Again the factor of $4$ comes from the Boltzmann equation for the relativistic $N$. At $T=T_N$, around $74\%$ of $N$ have decayed. }
\begin{equation}
\laq{TN}
T_{N}\sim 2\EV  \cdot C_{\rm decay}^{1/3}  \(\frac{M_N}{1\GEV}\)^{2}\(\frac{\theta}{10^{-6}}\)^{2/3}
\(\frac{10^{6}\GEV}{E^0_N}\)^{1/3}.
\end{equation} 
Here  $E^0_N$ is today's energy corresponding to $E_N$ if $N$ would not decay defined by
\beq
E^0_N\equiv \frac{E_N}{ (1+z)},
\eeq
where $z$ is the red-shift.
This satisfies
\beq
\laq{E1}
E^0_N\simeq 2\times 10^8\GEV  \(\frac{11}{g_{s\star}(T_R)}\)^{1/3} \(\frac{1\MEV }{T_R}\) \(\frac{m_\f}{M_{\rm pl}}\),
\eeq
where $g_{s \star}$ counts the relativistic degrees of freedom for the entropy density.
If we require $m_\f\lesssim 10^{19}\GEV$,  and $T_R \gtrsim \O(1)\MEV$ for the BBN bound~\cite{Kawasaki:1999na,
Kawasaki:2000en, 
Hannestad:2004px, Ichikawa:2006vm, DeBernardis:2008zz, deSalas:2015glj, Hasegawa:2019jsa}, the maximum energy is predicted to be $10^8-10^9\GEV.$ 
This will be the typical energy for the experimental events. It is noteworthy that the energy carries the information on the ratio $m_\f/T_R.$

The condition that $N$ decays before today can be obtained by solving \linebreak $\G_N^{\rm decay}[E_N^0]\gtrsim (13.8{\rm Gyr})^{-1}$. This yields,
\beq
\laq{decaycons}
E_N^0\lesssim E_N^{0\,{\rm max}}\equiv 10^8\GEV\, C_{\rm decay}  \(\frac{ M_N}{0.1\GEV}\)^6 \(\frac{\theta}{10^{-7}}\)^2,
\eeq
for $E_N\gg M_N,$ and $M_N\ll 100\GEV.$ We will see below that, if the decay happens at late enough time, the active neutrinos travel until today. In the next section~\ref{sec:H0}, we will study the phenomenology with $\G_N^{\rm decay}[E_N^0]\lesssim (13.8{\rm Gyr})^{-1}$, i.e. most of the $N$, produced during the reheating, survive until today.

\paragraph{Survival of the signal II: No scattering of $\nu$}

The $N$ decay produces active neutrinos at an $\O(1)$ branching fraction. The active neutrino today carries the kinetic energy 
\beq 
\laq{E2}
E_\nu =\O(0.1-1)\times E^0_N.
\eeq 
If it interacts frequently with the thermal bath, 
there would be a cascade of SM interactions including weak interactions. The interactions set a cutoff for $E_\nu$ as a function of the time scale for $N$ decays.   
Turned around, if the neutrino is produced late enough or with low enough energy it can reach Earth freely. 

When the decay takes place after electron -- positron annihilation, the neutrino mostly interacts with the cosmic neutrino background. The cross section of the neutrino-(anti)neutrino scattering is given as 
\begin{equation}
\s_{\n\n}= \frac{C_{\rm scat}}{\pi} G_F^2 E_{\rm cm}^2 ~~(E_{\rm cm} \ll m_Z)
\end{equation}
where $C_{\rm scat}=\O(1-10)$ {accounts for uncertainties inherent in the hadronic processes. The} center-of-mass energy is $ E_{\rm cm}^2\sim 2E_\nu \max[T_\nu, m_\nu],$ with the cosmic neutrino background temperature $T_\nu\approx (4/11)^{1/3}T$.\footnote{\label{ft1}We do not consider processes with $E_{\rm cm}\gtrsim m_Z$ since the required energy is greater than the cutoff energy from the non-resonant scattering. 
However the processes become important if the late decay of $N$ produce very energetic neutrinos. This does not happen in the parameter range, $T_R\gtrsim 1\MEV, m_\phi \lesssim 10^{19}\GEV$. }
The scattering rate of the neutrino is given as 
\begin{equation}
\laq{rate1}
\G_{\n}^{\rm scat} \sim \s_{\n\n}\times n_\nu\sim C_{\rm scat}G_F^2 E_{\rm cm}^2 \frac{T_\nu^3}{\pi^3}.
\end{equation}
The Universe is opaque to neutrinos if $\G_{\nu}^{\rm scat} \gtrsim 4H$ after the decay of $N$. Since $\G_{\nu}$ decreases faster than $H$, it is enough to compare the rates at the time when $N$ decays. 
This sets a cutoff energy of $E_\nu$ in terms of $T_N$
\beq
\laq{cutoff}
E_\nu \lesssim  10^8\GEV \, C_{\rm scat}^{-1} \(\frac{1\EV }{T_N}\)^3.
\eeq
By inserting $T_N$ from \Eq{TN} and taking $E_N^0\sim 10\, E_\nu$, one can derive a condition for $\nu$ to reach Earth,
\beq
\laq{nuloss}
\theta\lesssim 10^{-5} \(C_{\rm decay} C_{\rm scat}\)^{-1/2}\( \frac{1\GEV}{M_N}\)^{3}.
\eeq
This is independent of $E_N^0.$ 
This condition is justified when $N$ decays in the radiation dominated era. However if we take $m_\phi \lesssim 10^{19}\GEV$ and $T_R \gtrsim 1\MEV$ this is nearly automatically fulfilled since \Eq{E1} almost fulfills the inequality \eq{cutoff} with $T_N=1\EV$.\footnote{We can consider the late-time scalar decays to $NN$, in which case we may have $E_N^0\gg 10^9\GEV.$ In this late time decay scenario, the bound on $\theta$ can be alleviated by an order of magnitude for $E_N^0\ll 10^{14}\GEV$. When $E_N^0\sim 10^{14}\GEV$, however, there is no parameter region left due to the $Z$-resonance effect mentioned in footnote.\,\ref{ft1}.   }

\paragraph{Primordial neutrinos reaching Earth}
In Fig.\,\ref{fig:para}, we show the parameter region where the active neutrinos originating from the inflaton decays can reach Earth. 
This summarizes the following conditions:
rare energy loss of $N$ \eq{th2} [below blue dashed lines for $T_R=1\MEV$ and $100\MEV$], $N$ decays \eq{decaycons} [$E^0_N$ smaller than the contour value], 
rare scattering of produced $\nu$ \eq{nuloss} [left to red shaded region]. Here we have fixed $C_{\rm scat}=1 \AND C_{\rm decay}=3.$

We also show the see-saw relation by using the mass scales from neutrino oscillation experiments as the purple band corresponding to $\theta^2 M_N= [0.009-0.05]\EV.$ 
We recall that one can go above and below the purple band by considering approximate lepton number conservation (by introducing more than 1 flavor) and the possibility that $N$ gives the mass to the lightest active neutrino which can be lighter than the mass scale indicated by neutrino oscillations, respectively. Also, even if $E_N^0\gtrsim E_N^{0{\rm max}},$ we can have active neutrinos from a decaying $N$ (see sec.\,\ref{sec:HBSN}.). 

Furthermore we adopt the accelerator bounds from~\cite{Beacham:2019nyx,deGouvea:2015euy,NA62:2020mcv} for the case $\theta=\theta_e$ as a comparison [green shaded region], whereas the future reach~\cite{Beacham:2019nyx,Adams:2013qkq} of various experiments are shown as colored lines (see figure caption for details on the experimental data). We also note that when $M_N\sim \O(100)\MEV$, the $N$ enhances the neutrinoless double beta decay rate~\cite{Benes:2005hn,Atre:2009rg,Deppisch:2015qwa,deGouvea:2015euy} (not shown in the figure). Although the rate calculations, especially nuclear matrix elements, suffer from large uncertainties, this may give a strong hint for the scenario when the decay is discovered in the future. 
We notice again that there is an implicit constraint  of $E_N^{0} \lesssim 10^{8-9}\GEV$ from $m_\f\lesssim 10^{19}\GEV$ and $T_R\gtrsim 1\MEV$.

Consequently, we find that the active neutrinos 
originating from the inflaton decays can reach Earth without losing their energy by ways other than the red-shift. 
This is a central result of our paper. 

Finally, let us make two comments. 
First, right-handed neutrinos with momenta around $T$ are not produced significantly from the scattering between the thermal plasma since the constraint \eq{th2} restricts the thermal production rate.
Thus the bounds from the BBN and CMB for the low energy modes of $N$ given, e.g., in Ref.\,\cite{Ruchayskiy:2012si} are highly alleviated and we do not consider them in this paper. 
Second, although we have focused on $\f$ as an inflaton or a modulus, we can also consider $\f$ as any scalar or fermion that is non-relativistic and decays at the temperature $T=T_R.$
Our conclusion does not change even if $\f$ does not dominate the Universe, in which case we could have a decay temperature smaller than $\MEV$.

\section{Probing primordial neutrinos}
\label{sec:3}
Let us consider the active neutrino flux originating from the $\f$ decays. 

As a first step, we can consider
the differential flux of $N$ with $y=0$, i.e. without decays. By taking into account the red-shift, this is given by
\begin{eqnarray}
\frac{d^2\Phi_N}{d \Omega d{E_N^0}} \,=\, 
 \frac{\Gamma_{\phi\rightarrow NN}} {4\pi}   \int_{t_{R}}^{t_0}{dt'  r^{\rm scat}_N n_\phi(t')(1+z)^{-3} \frac{ d E_N'}{d E_N^0 } {J_N(E_N')}.}
\end{eqnarray}
Here, $\log{r^{\rm scat}_N}\sim -\Gamma_N^{\rm scat}/4H|_{T=T_R}$ as a function of $E_N'$ quantifies the survival probability from $N$-scattering with the thermal bath ($r_N^{\rm scat}=1$ for $y=0$), 
 and $E'_N=(1+z)E^0_N$ is the $N$ energy just after the decay. 
$(1+z)^{-3}$ represents the dilution of the flux due to the cosmic expansion. 
  The decay rate of $\f$ to $NN $ is related to the total decay width $\G_\f$ via
 \begin{equation}
 \Gamma_{\phi\rightarrow NN}=B_{\f\to NN}\G_\f,
 \end{equation}
 where $B_{\f\to NN}$ is defined as the branching ratio.  The spectrum of the monochromatic $N$ produced by a $\f$ decay is given by
\begin{eqnarray}
J_N(E)\,\equiv\, 
2 \delta{(E-\frac{m_\phi}{2})} ~.
\end{eqnarray}
The number density of $\f$ at the cosmic time $t$ is,
\begin{eqnarray}
n_\phi(t) \,\equiv\, r_\f \frac{ g_\star[T_\f] \pi^2 T_{\f}^4}{30 m_\phi } \frac{(1+z)^3}{(1+z_{\rm \f})^3}e^{-t \G_\f} ~.
\end{eqnarray}
Here $z_{\rm \f}$ is the red-shift at the time when $\f$ decays and $T=T_\f$ is the corresponding temperature defined in \Eq{reht}.
$r_\f$ represents the fraction of the energy density of $\f$ of the total energy of the Universe just before the decay. 
If $\f$ dominates the Universe $r_\f \approx 1$, we can identify $T_\f$ as the reheating temperature
\begin{equation}
T_R= T_{ \f}~~~ {[{\rm if}~ r_\f \approx 1]}.
\end{equation}
$r_\f \approx 1$ corresponds to the case that we have focused on so far, i.e. where $\f$ is the inflaton. 
However, as already mentioned our analysis also applies to the case that $\f$ is a non-relativistic subdominant component of Universe with  $r_\f\ll1$ and  $T_R$ replaced with $T_\f$.

The red-shift parameter $z$ is related to cosmic time $t$ by
\begin{eqnarray}
\laq{H2}
\frac{dt}{dz} \,= \frac{1}{(1+z) H}. 
\end{eqnarray}
At small red shift $H$ is taken as  $H=H_0 \sqrt{\Omega_M (1+z)^3+\Omega_r (1+z)^4 +\Omega_\Lambda}$ to take into account the effect from matter and dark energy.  
$H_0\simeq 1.4\times 10^{-42}\GEV$ is the present Hubble constant, and $\Omega_{M} \simeq 0.32, ~\Omega_\L\simeq 0.68, \AND \Omega_r\simeq 9.2\times10^{-5}$ denote the density parameter of matter, the cosmological constant and radiation, respectively~\cite{Aghanim:2018eyx}. For $T> 1 \KEV$, the Hubble parameter is obtained from the previous definition $H=\sqrt{g_\star\pi^2 T^4/90M_{\rm pl}^2}$ so that  
the entropy release from the decoupling of the SM particles is taken into account.

In a next step we can convert the 
 $N$ flux, $\frac{d^2\Phi}{d \Omega d {E_N}}$ with $y=0$, into the neutrino flux with $y\neq 0$.  Let us consider, for simplicity, a situation where $N$ mostly mixes with the electron neutrino, 
$\theta\approx \theta_e$.\footnote{
This assumption is not crucial in deriving our conclusions including the shape of flux. Indeed, as we will see, the decay channel involving electrons provides only for a $\sim 10\%$ effect.}
We then have the three body decay process
\begin{equation}
N \to \nu XY
\end{equation}
where $XY= e^- e^+, \bar{\nu}\nu.$
For $1\MEV\ll M_N<140\MEV$, i.e. below the pion mass scale and above the electron mass scale, hadronic channels are kinematically forbidden and we obtain $C_{\rm decay}\approx 2.3$ for the three body processes.
The branching fraction to electrons in this case is $B_{N\to \nu e^- e^+}\approx 0.11,$ and all those decay channels involve a neutrino. 
By imposing the ``see-saw relation" $\theta^2 M_N= [0.009-0.05]\EV$, we find that the scenario can be consistent with $T_R\lesssim 10-100\MEV$ in Fig.\,\ref{fig:para}. Moreover, the reheating temperature can be higher if $N$ corresponds to the lightest active neutrino. 
The distribution of the neutrino energy can be estimated as, 
\begin{equation}
P[x]\approx-18.2x^2(x-1.21)
\end{equation}
in the rest frame of $N$, where $x= 2 E_\nu^{\rm rest}/M_N$.
We take this as the typical energy distribution of $\nu$ and neglect other corrections.
In the boosted frame, $E_N\gg M_N$, one obtains 
\begin{equation}
E_\nu \simeq E^0_N \frac{x}{2} (1+\cos\Theta),
\end{equation}
where $\Theta$ is the angle between the direction of $\nu$ and the boosted direction in the rest frame of $N$.
The neutrino flux is,
\begin{equation}
\frac{d^2\Phi_\nu}{d \Omega d {E_\nu}}= B_{N\to \nu XY}\int{ d E_N  r_{N}^{\rm decay}   \frac{d E^0_N}{d E_\nu} \frac{d^2\Phi_N}{d \Omega d {E^0_N}}},
\end{equation}
where
\begin{equation}
 \frac{d E_N^0}{d E_\nu}=\int_{-1}^{1}{{d\cos[\Theta]}\int_{0}^{1}{dx \frac{1}{2} \delta\(E_\nu -\frac{x}{2}(1+\cos[\Theta])E_N^0 \) P[x]}},
\end{equation}
and $B_{N\to \nu XY}$ is the decay branching ratio of $N$ to $\nu XY$, which is $\approx 1$ for our case, $r_{N}^{\rm decay}$ represents the fraction of $N$ that decays. 
It satisfies $\log{(1-  r_{N}^{\rm decay}  )}\sim -\G_{N}^{\rm decay}\times 13.8$~Gyr. 
Notice that we have neglected the scattering between a produced neutrino and the cosmic neutrino background. In the mass range $M_N<140\MEV$ this is justified from Fig.\,\ref{fig:para}.\footnote{In the presence of scattering we must take into account of the cascade of neutrinos which depends on the decay rate of $\G_N^{\rm decay}$. 
When $M_N$ is much larger than the cascade limit in Fig.\,\ref{fig:para},  the spectrum (cf., e.g., Ref\,\cite{Ema:2014ufa}) has a typical energy as a function of the timescale at which $N$ decays.}

\begin{figure}[t!]
\begin{center}  
\includegraphics[width=135mm]{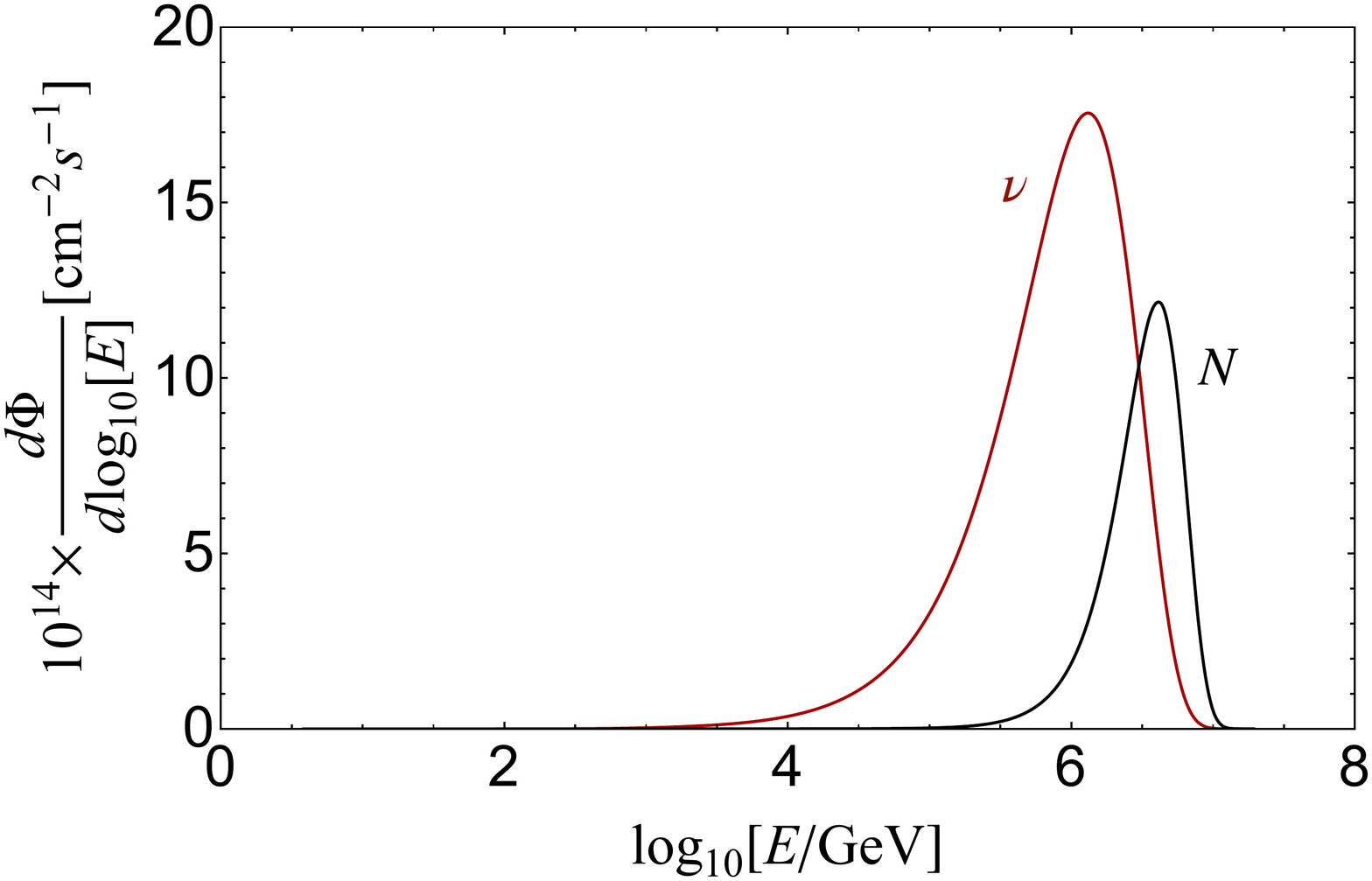}
\end{center}
\caption{The $N$ flux [black] by taking $y=0$ (i.e. the $N$ are not decayed) and $\nu$ flux [red] from $\f$ decays.  
 Here $m_\f=6\times 10^{17}\GEV$, $T_R=10\MEV$.  $r_\f r_N^{\rm scat} B_{\F\to NN}=10^{-7}$ [black] and $r_\f r_N^{\rm scat} r_N^{\rm decay} B_{\F\to NN} B_{N\to \nu XY} 
 =10^{-7}$ [red] are taken.} 
\label{fig:fl}
\end{figure}

In Fig.~\ref{fig:fl}, we show the differential flux of $\nu$ [red] and $N$ [black] from $\f$ decays for a simple benchmark point.
As the red line corresponds to the daughter $\nu$ of the $N$ due to the $N\to\nu XY$ decay it peaks at somewhat lower energies.
The larger area of the red curve arises from the fact that $XY$ can be $\bar{\nu}\nu$ and therefore on average more than one  neutrino  is produced in each decay.

\begin{figure}[t!]
\begin{center}  
\includegraphics[width=135mm]{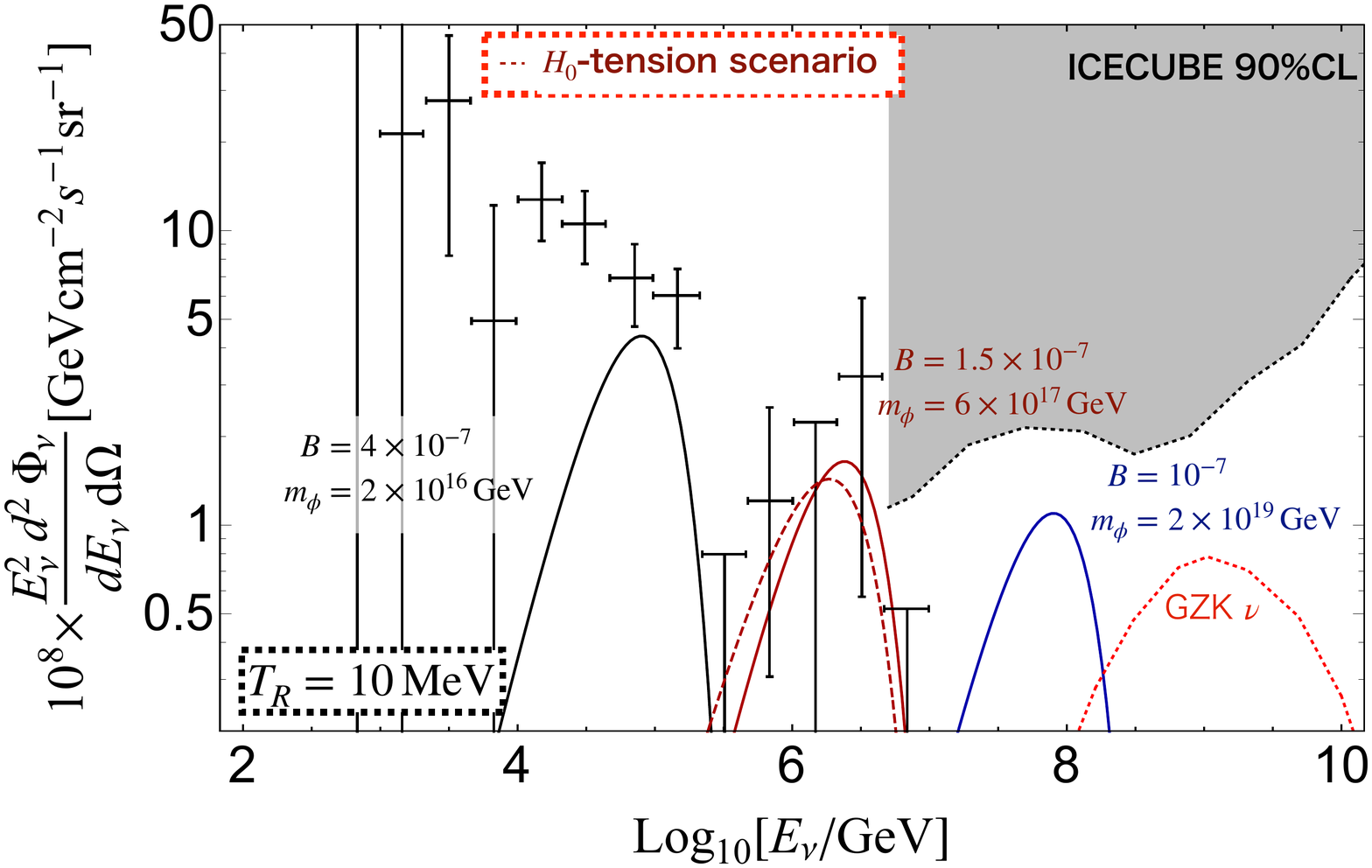}
\end{center}
\caption{Primordial neutrino flux. $m_\f=[2\times 10^{16},6\times 10^{17},2\times 10^{19}]\GEV$ and $B=[4, 1.5, 1]\times 10^{-7}$ for black, red, and blue lines from left to right, respectively. 
The red dashed line may be consistent with alleviating the $H_0$-tension (see Sec.\,\ref{sec:HBSN}), where $r_\f r_{N}^{\rm scat}  B_{N\to \nu XY} B_{\f\to NN}=0.1, m_\f=6\times 10^{17}, M_N=7.5\MEV, \theta=4\times 10^{-8}, C_{\rm decay}=2.3$. In this case only a small fraction of the $N$ have decayed by today. The spectra do not change much when altering $T_R$ while keeping $m_\phi/T_R$ and $B$ fixed.
The error bars are data points and upper limits observed at $1\sigma$~\cite{Aartsen:2017mau}. The gray shaded region may be excluded at the $2\s$ level~\cite{Aartsen:2018vtx}. 
For comparison we also show a possible GZK flux taken from Ref.~\cite{Ahlers:2012rz}.
} 
\label{fig:2}
\end{figure}

In Fig.~\ref{fig:2}, we depict the predicted neutrino flux originating from the $\f$ decays with $T_R=10\MEV, r_\f=1$ and different masses.
In this part we ignore the energy dependence of $r_N^{\rm scat}$ and $r_N^{\rm decay}$ and take 
\begin{equation}
    r_N^{\rm scat}, r_N^{\rm decay} \approx 1.
\end{equation} 
This is justified as long as the parameter values are away from the boundaries of the constraints in Fig.\,\ref{fig:para}.
For convenience we introduce the quantity,
\begin{equation}
\laq{Bdef} 
B\equiv r_\f  r_{N}^{\rm scat} r_N^{\rm decay} B_{\f\to NN} B_{ N\to \nu XY},
\end{equation}
which represents the amount of the neutrino density originating from $\f$ decays. 
The solid lines are for $m_\f=[2\times 10^{16},6\times 10^{17},2\times 10^{19}]\GEV$ and $B=[4, 1.5, 1]\times 10^{-7}$ from left to right. 
The red dashed line indicates a parameter choice suitable to explain the $H_{0}$-tension as discussed in Sec.~\ref{sec:HBSN}. For comparison the red  dotted line gives a prediction for the expected flux of Greisen-Zatsepin-Kuzmin (GZK) neutrinos from~\cite{Ahlers:2012rz}. We can see that this is still comparable or even a  bit smaller, allowing for a relatively clean signal from our messenger neutrinos from reheating.
The constraints from the IceCube experiment~\cite{Aartsen:2017mau, Aartsen:2018vtx} are shown in grey and the crosses denote further measured data points. 
We find that for $1\TEV<E_\nu<10^9\GEV$ the scenario can be consistent with the current observation and limits on the neutrino flux as long as $B\lesssim 10^{-7}-10^{-8}$. 
Moreover, the data in the PeV region could perhaps even be explained by the ``primordial" neutrino as a remnant of the reheating. 

For $m_\f\sim 10^{19}\GEV$ with $T_R\sim \O(1-10)\MEV$,
 the neutrino energy observed can be in a range similar to that of the GZK neutrinos~\cite{Greisen:1966jv,
Zatsepin:1966jv}.

Note again that in order to have the energetic primordial neutrinos today, 
we should introduce the right-handed neutrinos for two reasons.
For us the main reason is that the energetic (active) left-handed neutrino has a suppressed mean-free path just after the reheating, and
 the mixing between right-handed and left-handed neutrinos can lead to a longer one. 
Secondly, we need the inflaton to be heavy and to decay into very energetic neutrinos. For them to still be sufficiently energetic after a long period of red-shifting, we need a suitable UV completion to generate the neutrino masses. 
In this sense this is a relatively minimal model leading to primordial neutrinos reaching Earth. It predicts the typical primordial neutrino spectra corresponding to the two-step process $\f \to N N\to \nu XY.$ 
Future observation of the cosmic-ray neutrino spectrum (especially around PeV events) can test this scenario and thus the reheating.

\paragraph{Other constraints}
Let us mention other relevant constraints on the $N$ decays. As already explained, if $M_N\gtrsim 1\MEV$ and the mixing to the electron neutrino is non-vanishing, $N$ can decay to electrons. 
The injected energetic electrons interact with the ambient plasma, which is constrained from the CMB and BBN. 
The amount of energetic electrons from the decay of $N$ can be characterized by the quantity
\beq
B_e=r_\f   
\times B_{\f\to NN}\times B_{N\to e^+ e^-\nu},
\eeq 
where $B_{N\to e^+e^{-}\nu}$ is the $N$ decay branching fraction to electrons and we again have taken \linebreak $r_N^{\rm scat}r_N^{\rm decay}=1.$

\begin{figure}[t!]
\begin{center}  
\includegraphics[width=135mm]{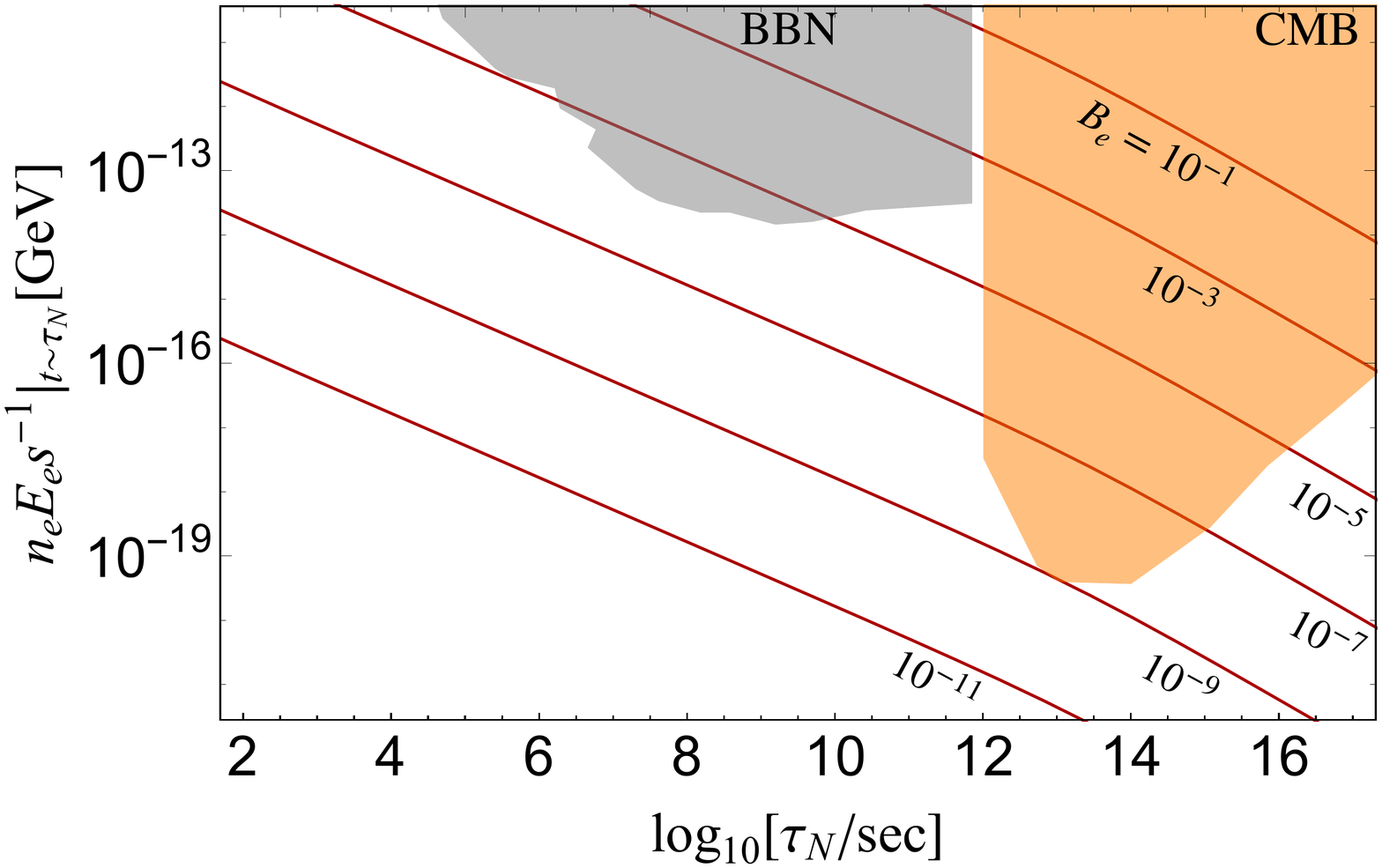}
\end{center}
\caption{The constraint on the decays $N\to \nu e^-e^+$. The horizontal axis indicates the life-time of $N$, and the vertical axis gives the energy density to entropy density for the electron produced by the $N$ decay. The red lines show the predictions with $B_e=10^{-1},10^{-3},10^{-5},10^{-7}, 10^{-9}, 10^{-11}$ from top to bottom with $m_\phi=10^{17}\GEV, T_R =10\MEV$, but the predictions are not sensitive to the values of $m_\phi$ and $T_R$.  The gray (orange) region is excluded due to the BBN bounds~\cite{Kawasaki:2017bqm} (CMB bounds~\cite{Poulin:2016anj}).}\label{fig:3}
\end{figure}

The constraint is shown in Fig.\,\ref{fig:3} in the plane of the $N$ life-time, $\tau_N\equiv 1/\Gamma^{\rm decay}_N$, (including the boost factor) vs the ratio of energy density of the electrons produced at the $N$ decays to entropy density. 
 The CMB and BBN constraints are derived from the references~\cite{Poulin:2016anj} and~\cite{Kawasaki:2017bqm}, respectively. 
The references give constraints to the energy to entropy ratio of a non-relativistic particles decaying into electron--positron pairs. 
This is different from our relativistic $N$ decays.  However, we can directly adopt it for the constraint on $n_e E_e/s|_{t\sim\tau_N}$ as this exhibits the same scaling of the injected energy density.
In the references, the injected energy per time at $t\ll 1/\Gamma_{\rm decay}$ is $\rho_{\rm mother}\Gamma_{\rm decay}$ which scales as $a^{-3}$ since the decay rate $\G_{\rm decay}$ is constant.  Here $\rho_{\rm mother}$ is the 
energy density of the non-relativistic particle. 
On the other hand, in our case, the injected energy per time at $t\ll \tau_N$ is $\rho_{N} B_{N\to e^+e^{-}\nu} \G_{N}^{\rm decay}\propto a^{-4} \times a=a^{-3}$. Here we have used the decay rate \eq{decayrate} scaling 
$\G^{\rm decay}_N\propto a$. With $t\gtrsim \tau_N$ the $N$ energy is transferred quickly into $\nu$ and $e$ as in the non-relativistic case. 
Thus the constraints from the references should be useful for us.  
To derive the contours we used that on average $\simeq 11/20$ of total energy of $N$ is transferred into the electrons when $N\to e^+e^-\nu$ happens. 
One finds that with $B_e\lesssim 0.001$, there is a viable parameter region depending on the life-time of $N$ in the time scale of our interest. This constraint may be avoided if $|\theta|\gg |\theta_e|$ or if $M_N\lesssim 1\MEV$ to forbid the channel kinematically.

$N$ or the produced neutrinos also contribute to the dark radiation which is represented as the deviation from the effective neutrino number
  \begin{equation}
 \D N_{\rm eff}
 \sim 6 r_\f r_N^{\rm scat} B_{\f\to NN} \(\frac{11}{{g_{s\star}(T_R)}^{4}g_\star(T_R)^{-3}} \)^{1/3} \laq{Neff}.
 \end{equation}
Thus as long as $r_\f \times B_{\f \to NN}\times r_N^{\rm scat}$ is much smaller than $1$ we do not have a significant contribution to the $\Delta N_{\rm eff},$ e.g. $|\D N_{\rm eff}|<\O(1)$ from a BBN bound as given, e.g., in Ref.\,~\cite{Kawasaki:2017bqm} (the implications for the CMB of $\D N_{\rm eff}$ will be discussed later).

However, if $\Delta N_{\rm eff}$ is not too small, a constraint from the late time scattering of $N$ becomes relevant. Even if, according to Eq.~\eq{th2}, only a fraction of the $N$ scatter, this can induce noticeable effects. The scattering produces charged leptons as in Fig.~\ref{fig:diag1}. These energetic leptons inject energy similar to the electrons originating from $N$ decays. 
We estimate the produced energy density of energetic electrons at $t$ by $n_e E_e \sim \rho_N \Gamma_N^{\rm scat}/H$.  We impose $n_e E_e$ at any $t$ is smaller than the upper-bounds in Fig.~\ref{fig:3} (again using the data from~\cite{Kawasaki:2017bqm,Poulin:2016anj}) at any $t$ before the decays of $N$ and if $E_{\rm cm}\gg m_{Z,W}.$
With this we obtain 
\beq
\laq{scatCMB}
\theta \lesssim (10^{-6}-10^{-5})\sqrt{\frac{0.1}{\Delta N_{\rm eff}}}
\eeq
with $T_N\lesssim \KEV.$ For instance, for $\theta=10^{-5}-10^{-4}$, $\Delta N_{\rm eff} \lesssim 10^{-2}-10^{-1}$ may be satisfied. 

\begin{figure}[t!]
\begin{center}  
\includegraphics[width=135mm]{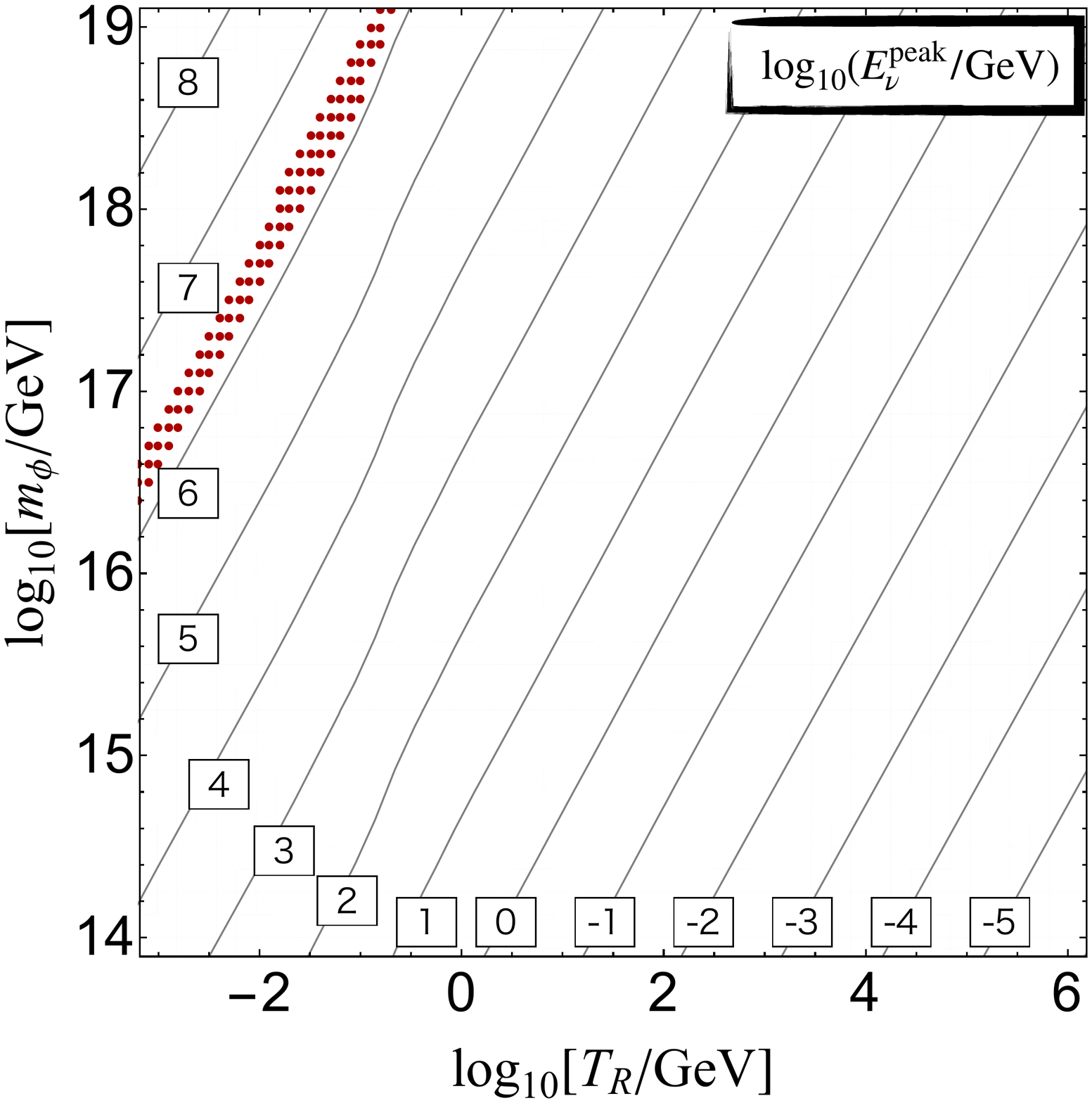}
\end{center}
\caption{The contours of the peak energy  $\log_{10}(E^{\rm peak}_\nu/\GEV)$ of $E_\nu^2 d \F_\nu/d E_\nu$ in $m_\f-T_R$ plane.   
Red points can explain the PeV events at the 1$\s$ level observed at IceCube.  } 
\label{fig:epeak}
\end{figure}

\begin{figure}[t!]
\begin{center}  
\includegraphics[width=135mm]{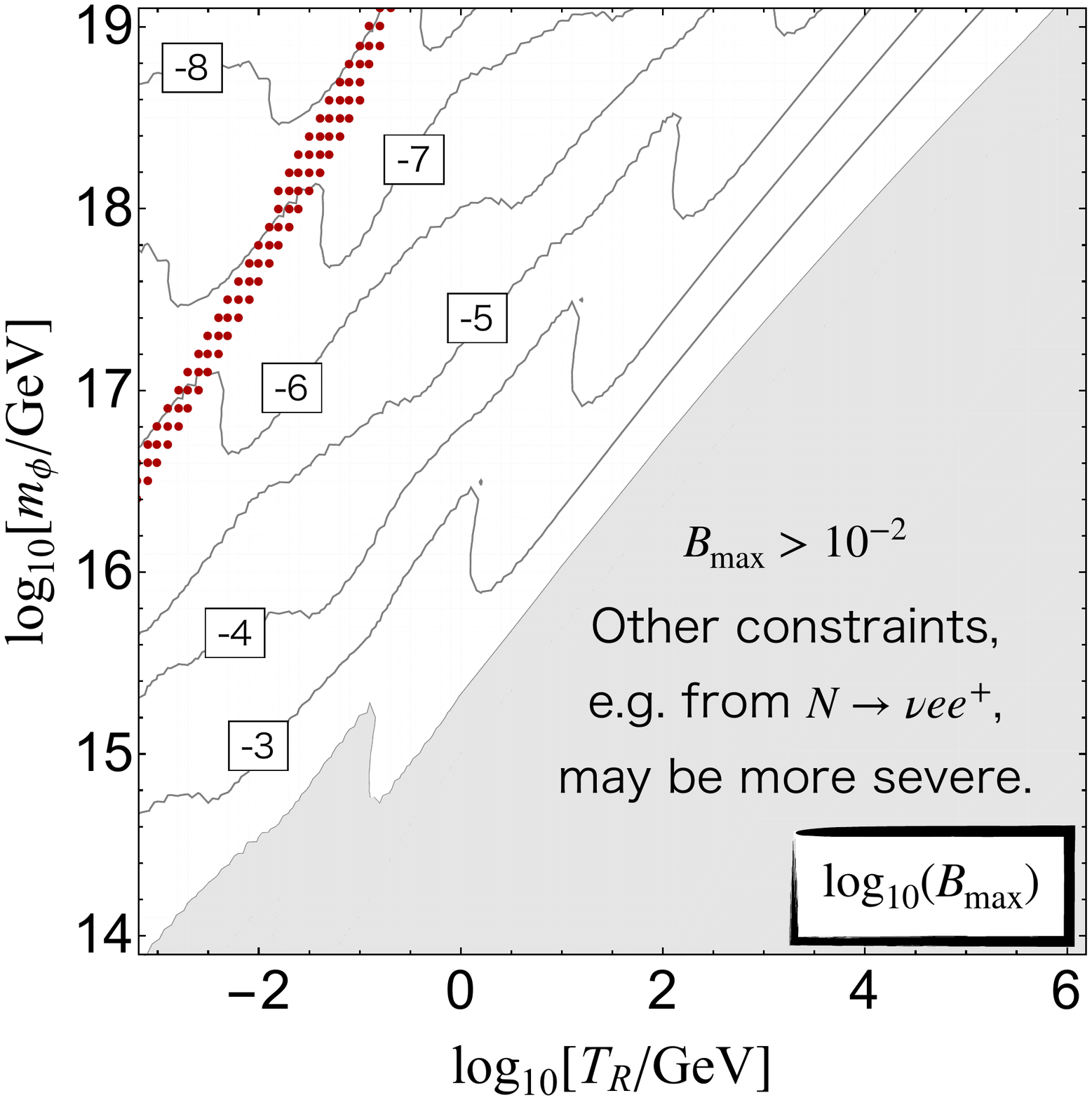}
\end{center}
\caption{Contours of $\log_{10}(B_{\rm max})$. The gray region has $B_{\rm max}>10^{-2}$ from the constraints to neutrino flux. However, this may not be  consistent with the other flux constraints in $N$ decays which, at least for mixing with the electron neutrino require $B<10^{-2}$, cf. Fig.\,\ref{fig:3}.  } 
\label{fig:Bmax}
\end{figure}

\paragraph{Parameter region for reheating}
We have performed a numerical analysis to clarify the parameter region of the inflaton sector in terms from the observables of the active neutrino. 
In Fig.~\ref{fig:epeak}, we show the contours of the peak energy of the neutrino spectrum, $E_\nu^2 d \F_\nu/d E_\nu$. The red points represent the region that can explain the PeV events at the 1$\s$ level.
In Fig.~\ref{fig:Bmax} we show the upper limit of \Eq{Bdef} from the IceCube bound and that $d \F_\nu/d E_\nu$ is smaller than the atmospheric neutrino flux~\cite{Katz:2011ke} in the range $0.1\GEV\leq E_\nu\leq 100\GEV$. We exclude the region with $B>10^{-2}$. There the CMB and BBN bounds given in Fig.~\ref{fig:3} are more severe (we stress, however, that they apply directly only in the case of mixing with the electron neutrino) and so does the constraint on the late-time scattering of $N$.
Consistent with the ``see-saw relation'' $\theta^2 M_N= [0.009-0.05]\EV$, the PeV events can be explained by the primordial neutrinos if $m_\f=\O(10^{16-19})\GEV$ and $T_R=\O(1\MEV-1\GEV)$.\footnote{In the parameter region we have focused on, $\f\to NN\to \nu+\cdots$ is suppressed compared to the decays to $hh$.  For this we need to fulfill the condition $\l \sim \sqrt{B_{\f\to NN}} A/m_\f\ll A/m_\f$. 
Such a hierarchy may be due to a (baryon minus) lepton number symmetry which is recovered with $\l, M_N\to 0$. In the next section we see one can have $\l \sim A/m_\f$ to have  $N$ as a relativistic dark particle and explain the PeV events. For other symmetry arguments to explain the fine-tuning, see Appendix~\ref{ap1}.}

\section{\texorpdfstring{$H_0$}{H0} tension, PeV events, and relativistic dark particles from reheating} \label{sec:H0}
 
So far we have focused on active neutrinos produced in the decays of $N$ as a possible observable. However, the possibility of messengers from inflation can apply to more general situations. A first option is that the $N$ do not decay and therefore contribute a highly relativistic DM component (one could also consider this as dark radiation). 
Beyond that we can of course also consider the more general situation of a dark particle being produced in connection with reheating that can serve as a messenger from this era. We will henceforth refer to all these as relativistic dark particles (RDP).

Let us start with a heavy $\f$ decaying to an RDP. Such an RDP may be tested from ground-based experiments such as direct detection experiments and neutrino observatories (see Ref\,\cite{Arguelles:2019xgp} and the references therein).  
 This then also gives a probe of the reheating.  It may be interesting to 
notice that if the RDP contributes to $\D N_{\rm eff}=\O(0.1)$, the 
$H_0$-tension can be alleviated~\cite{Mortsell:2018mfj, Riess:2019cxk}, and also can be tested 
from precisely measuring the dark radiation from the CMB and baryon acoustic oscillation experiments in the future~\cite{Kogut:2011xw, Abazajian:2016yjj, Baumann:2017lmt}. 
The facts that the RDPs can have sizable contributions to the energy density of the Universe and the spectra are  
 distinguishing features compared to the highly-boosted DM generated by SM energetic cosmic-ray via DM-SM particle interactions~\cite{Yin:2018yjn,Bringmann:2018cvk, Ema:2018bih}. 
We will now see that we can test such RDPs also in ground-based experiments.

\subsection{Highly-boosted sterile neutrinos}
\label{sec:HBSN}
For concreteness, let us consider again the Lagrangian \eq{Lag} with $N$ and with $y$ or $M_N$ so small that the (boosted) $N$ is meta-stable, i.e. $\G_N^{\rm decay}< (13.8\,{\rm Gyr})^{-1}$. Let us now see how we can probe reheating in this case.

A typical flux of $N$ is given as the black line in Fig.\,\ref{fig:fl}. In the following, let us study the detection of the reheating through the sterile neutrino by considering the decays. 
In particular we consider that relativistic $N$ contribute significantly to the dark radiation with $\D N_{\rm eff}=\O(0.1),$ which is the case for $\l\sim A/m_\f, $
 and thus the $H_0$-tension is alleviated in the following scenarios.

A small but non-vanishing decay rate of $N$ leads to a decay probability $r_N^{\rm decay}\ll 1$.
One can estimate that 
\beq 
r_N^{\rm decay} \approx \int{dt \G_N^{\rm decay}[E_N^0(1+z)]}\approx 10^{-6} C_{\rm decay} \(\frac{M_N}{10\MEV}\)^6\(\frac{\theta}{3\times 10^{-8}}\)^2 \(\frac{5\times 10^6\GEV}{E_N^0}\)
\eeq
as a function of $E_N^0.$ 
 Even if $N$ composes a significant component of the dark radiation from \Eq{Neff}, one can obtain a small enough $B$ in \Eq{Bdef} (approximating $r_N^{\rm decay}$ as a constant) to be consistent with the experimental bounds. 
For instance, even with 
$r_N^{\rm scat} r_\f B_{\f\to NN}= \O(0.1)$  and thus $\Delta N_{\rm eff}= 
\O(0.1)$, this evades most of the experimental bounds. 
In particular, we take $\theta$ small enough to evade the bound \eq{scatCMB}. This makes it difficult to satisfy the ``see-saw relation", but $N$ may be the sterile neutrino responsible for the mass of the lightest active neutrino. 

In this case, the decay products, active neutrinos and electrons, are energetic, and would appear as cosmic-rays.  One can again constrain or test the scenario from the cosmic-ray neutrino 
searches. The difference from the previous section is that $N$ is decaying, which will cause a slight modification of the neutrino spectrum (see the $E_N^0$-dependence of $r_N^{\rm decay}$).

In particular, we may still explain the PeV events if $r_N^{\rm decay}=\O(10^{-6})$ given $B_{N\to \nu XY}\approx 1.$
The flux corresponding to this case is given as the red dashed line in Fig.\,\ref{fig:2} where $M_N=7.5\MEV, \theta=4\times 10^{-8},  r_N^{\rm scat} r_\f B_{N\to \nu XY}B_{\f\to NN}=0.1$ and $C_{\rm decay}=2.3$ (the estimated value in the SM when $N$ only mixes with $\nu_e$) are fixed. $m_\f=6\times 10^{17}\GEV, T_R=10\MEV$ are  same as the red solid line.  We find that the spectrum is slightly shifted to lower energies. This is because $N$ with lower energy have slightly larger decay rate due to the boost factor, and thus more $\nu$ are produced. That said, the spectrum is almost the same as the one from the 2 step decays in the previous section.  

\bigskip 

One may also wonder if direct scattering between $N$ and detector particles through mixing is relevant. In the detector, the ``effective'' neutrino flux\footnote{In principle one distinguishing feature from ordinary neutrinos would be that they have a very  long mean free path in medium. They would pass Earth essentially untouched while Earth is more or less opaque to energetic neutrinos with $E_\nu \gtrsim 10^{6}\GEV$ \cite{Nicolaidis:1996qu,Naumov:1998sf,Kwiecinski:1998yf,Giesel:2003hj} depending on the injection angle.}  is given by $ d \F_\nu^{\rm eff}/dE d\Omega \equiv \theta^2 d \F_N/dE d\Omega$. It has the shape of the flux of $N$, shown as the black line in Fig.\,\ref{fig:fl}, reduced by the mixing squared. Its magnitude is $\propto \Delta N_{\rm eff}\theta^2$. Taking into account the BBN and CMB constraints, Eq.~\eq{scatCMB}, the flux should be smaller than $ \O(10^{-13}) \GEV$cm$^{-2}$s$^{-1}$sr$^{-1}$ at an energy around $E_N^0$. This seems challenging to test in near future experiments.

\subsection{General RDPs stabilized by symmetry}
Up to now we have studied the flux, decay and detection of a very highly boosted very weakly interacting particle, concretely the right-handed neutrino, originating from inflaton decays. 
In particular the right-handed neutrino did not feature a conserved quantum number. Alternatively one could also imagine that the inflaton decays into particles that feature a stabilization by a symmetry and are therefore conserved. 
If the couplings to the SM particles are large enough at lower energy, it is not necessary to have the RDP highly energetic in order to test it.  
Then the reheating can be probed with a smaller $m_\f/T_R$ than the focused on in the previous sections.  

While it is tempting to also use this stable particle as dark matter, this is not necessary for most of the arguments given in the following.

Let us examine whether such an RDP can be a messenger from reheating. 
To this end, suppose that the scattering cross section is given in the form 
\beq 
\laq{scatDM} 
\s_{\rm scat}^{\rm RDP}= \sigma_{\rm today} (1+z)^{-n/2}
\eeq 
where $\s_{\rm today}$ is the scattering cross section per scatterer today. The red-shift dependence may come from the dependence of the center-of-mass energy 
$E_{\rm cm}\sim \sqrt{E m} \propto (1+z)^{1/2}\OR \sqrt{E T} \propto (1+z)$ at high enough $E_{\rm cm}$. In particular, $n$ is typically non-negative for renormalizable interactions at energies above the typical mediator masses. In general $n$ can depend on the energy regime (see Appendix \ref{ap2}). However, for simplicity, we only consider the case where $n$ is constant. We also assume throughout this part that the scattering of the RDP with energy $E_{\rm RDP}$ transfers the momentum of $\O(E_{\rm RDP})$ to the scatterer.
 
The more relevant quantity however, would be the scattering rate in a detector today, $z= 0$,
\begin{eqnarray}
\label{eq:rate}
\Gamma_{\rm detector}&\sim& n_{\rm RDP}\left(\frac{M_{\rm target}}{m_{\rm target}}\right)\sigma_{\rm today}
\\\nonumber
&\approx& 0.003\frac{\rm events}{\rm year}\left(\frac{M_{\rm target}}{1\,{\rm ton}}\right)\left(\frac{1\,{\rm GeV}}{m_{\rm target}}\right)\left(\frac{\Delta N_{\rm eff}}{0.1}\right)\left(\frac{1\,{\rm TeV}}{E_{\rm{RDP}}}\right)\left(\frac{\sigma_{\rm today}}{\rm pb}\right).
\end{eqnarray}
In the first line $n_{\rm RDP}$ is the density/flux of the RDP component today\footnote{We recall that for relativistic particle flux and density only differ by a factor of $c=1$.}  
\begin{equation}
n_{\rm RDP}\approx
18\,{\rm m}^{-2} s^{-1} \D N_{\rm eff}\left(\frac{1\,{\rm TeV}}{E_{\rm RDP}}\right),
\end{equation}
 $m_{\rm target}$ is the mass of the individual scatterer in the detector and $M_{\rm target}$ is the total mass of all the scatterers in the detector. 
This can be compared to the maximal allowed cross section for an RDP particle to reach the underground detector. The relevant cross section  (where the RDP scatters at most once on its way to the detector) is given by,
\begin{equation}
\sigma_{\rm max}=\frac{1}{{\rm{scatterers\,\,per\,\,area}}}=1.8\times 10^{7}{\rm{pb}}\left(\frac{1000\,{\rm ton}/{\rm m^2}}{\rm overburden}\right)\left(\frac{m_{\rm overburden}}{1\,{\rm GeV}}\right).
\end{equation}
where $m_{\rm overburden}$ is the typical mass of the individual scatterer in the overburden on the way to the detector. 
Below this cross section, the location of the experiment can be reached by the RDPs. 

Moreover we need to check whether the early universe is transparent to such an RDP produced at the reheating. 
To do this we can  
consider the scattering probability with the thermal bath,
\begin{equation}
\laq{Totint}
\int^{t_{\rm today}}_{t_{\rm start}} dt \, n(t)\sigma(E(t)).
\end{equation}
and ask if it is small enough. 
Here $n(t)$ is the number density of suitable the scatterers in the Universe. 
This then gives us a maximal cross section that can be seen in current/future DM/neutrino experiments.
We can now perform the integration,
\begin{eqnarray}
P_{\rm scatter}[t_{\rm start}]&=&\int^{t_{\rm today}}_{t_{\rm start}}dt n_{\rm today}\sigma_{\rm today} \left(1+z \right)^{3-\frac{n}{2}}.
\end{eqnarray}
This can be conveniently evaluated by using \Eq{H2}. 

For the Universe to be transparent from $t=t_{\rm start}$ to $t_{\rm today}$ we require in general, \linebreak $P_{\rm scatter}[t_{\rm start}]\lesssim 1$. However, if $\D N_{\rm eff}\sim \O(0.1-1)$, rare scatterings may transfer energy to the scatterers. This is constrained from BBN and the CMB~\cite{Kawasaki:2017bqm,Poulin:2016anj} as discussed around \Eq{scatCMB}. As a rough estimate this yields the limit $d P_{\rm scatter}[t]/dt \times 1/H \lesssim \O(10^{-6}-10^{-5})/\Delta N_{\rm eff}$ at $T \sim \KEV$. In the following examples, this na\"{i}ve constraint can be satisfied with $\Delta N_{\rm eff}=\O(0.1)$, i.e. while alleviating the $H_0$-tension. 

\paragraph{RDP coupled to baryons}
Let us now get a numerical estimate. For $n=0$, i.e. an energy independent cross section and using scattering on baryons only, $n_{\rm today}= 9.2\times 10^{-11}s_0$
with $s_0\approx(0.28 {\rm \,meV})^3$~\cite{Aghanim:2018eyx} being the entropy density today, 
 we find,
$
P_{\rm scatter}\approx 6.8\times 10^{-10} \left(\frac{\sigma_{\rm today}}{\rm pb}\right) 
$
at matter-radiation equality, which allows us ample room to have detectable cross sections according to Eq.~\eqref{eq:rate}.
A stronger constraint arises from earlier times, 
\begin{equation}
P_{\rm scatter}\approx 2.4\times 10^{-3} \left(\frac{\sigma_{\rm today}}{\rm pb}\right) \(\frac{T}{ 1\MEV}\)\(\frac{g^{-1/2}_\star g_{s\star} }{3.2}\) ~~({\rm if~ }n=0),
\end{equation} 
where $T$ is the temperature of the thermal bath at $t=t_{\rm start}$.

Therefore with $T_R=\O(10^{-3}-10^{-1})\GEV$ one can expect that the Universe is transparent to the RDP.
With $n>0$, $P_{\rm scatter}$ is smaller than this, e.g. with $n=1$ $P_{\rm scatter}\approx 6.1\times 10^{-8}  \(\frac{T}{ 1\MEV}\)^{1/2}\(\frac{\sqrt{g_\star}}{3.3}\)^{1/2} $. 
As a result, we can conclude that the RDP can travel to Earth if $n\geq 0$, which may be the case if the effective model describing the RDP interaction with the other particles is renormalizable (dimension$\leq 4$).\footnote{In many situations this will require inclusion of an extra mediator particle that is relatively light, possibly leading to further constraints on the model. See Appendix~\ref{ap2} for the case of a coupling with leptons.}

$\G_{\rm detector}$ can be enhanced with small $E_{\rm RDP}$ for a given $\D N_{\rm eff}$. For instance in DM direct detection experiments with $M_{\rm target}={\cal O}(1)\,$ton, $\O(1)\MEV \lesssim E_{\rm RDP}\lesssim \O(1)\GEV$, and $\s_{\rm today}\sim $pb,\footnote{The dark particle may be produced accompanied by a mono-photon or mono-jet in hadron colliders. As a very naive first estimate we take the production cross section to roughly be $\sigma_{\rm monophoton/jet}\sim 1/(16\pi^2) \times \sigma_{\rm today}$ with a similar collision energy of $E_{\rm RDP}.$
$\s_{\rm monophoton/jet}$ should be below $\O(1)$ fb from collider experiments. This may set $\sigma_{\rm today}\lesssim 1\,$pb. However, the constraint is UV model-dependent. 
If the cross section increases with center-of-mass energy, the constraint on $\sigma_{\rm today}$ from LHC is more severe \cite{Khachatryan:2014rra, Aaboud:2017phn,Sirunyan:2017ewk}.}the event rate can be obtained as  $\O(1-10)\lesssim \G_{\rm detector} \lesssim \O(10^3-10^4)$ per year.  The recoil energy of a nucleon is $\O(1)\KEV \times (E_{\rm RDP}/1\MEV)^2$ which is in principle testable in direct detection experiments~\cite{Agnese:2015nto,Akerib:2016vxi,Tan:2016zwf,Angloher:2015ewa, Amole:2017dex}. Thus, the DM direct detection for the nucleon recoils may probe reheating with $10^{10} \lesssim m_\phi/T_R\lesssim 10^{13}.$

When $E_{\rm RDP}\gg 1\GEV$, on the other hand, the flux of the RDP is suppressed. 
However, it  may be tested in large volume detectors such as IceCube. 
Deep inelastic scattering similar to neutral current interactions induces shower- or cascade-like events. This would therefore be similar to one of the main channels for searching neutrinos. 
The events for neutral current interactions are more localized and spherical 
than those for a charged-current interaction of $\nu_\mu$, since the produced secondary particles soon decay or scatter in the medium (see Ref.\,\cite{Ahlers:2018mkf}).  
Therefore our scenario predicts spectra with a peak, around which larger ratio of the events of cascade-like to track-like than a neutrino produces will be seen. The peak energy again corresponds to the ratio of the inflaton mass to the reheating temperature. 
Since IceCube may test $E_{\rm RDP}\gtrsim 100\GEV$, reheating with $m_\f/T_R\gtrsim 10^{14}$ may be probed.

\paragraph{RDP coupled to leptons}

If the RDP dominantly interacts with the other particles, such as electrons, $P_{\rm scatter}<1$ becomes more difficult to be satisfied with $T_R\gtrsim m_e$, since the number density of scatterers is not suppressed by the baryon asymmetry. 
The integral \eq{Totint} is dominated by the period before the electron-positron annihilation (in terms of temperature), 
\beq
\laq{lepton}
P_{\rm scatter}\sim \int_{m_e}^{T}{ \sigma_{\rm today} \frac{d T'}{s_0 H } \frac{d s}{d T'}\(\frac{s}{s_0}\)^{-n/6-1/3} \frac{T'^3}{\pi^2} }.
\eeq
Here, we take into account $n(t)\sim T^3/\pi^2 \,\,(T\gg m_e)$, $\sim 10^{-10} T^3/\pi^2 (T\ll m_e)$, and $s(T')$ ($s_0$) is the entropy density at  $T=T'$ (today). 
In this case we need $n\geq 2$ to have the RDP travel freely with $T>1\MEV$ where $P_{\rm scatter}\sim (10^{-4}-10^{-3})(\sigma_{\rm today}/1\,{\textrm{pb}})$ independent of $T'$. 
One can understand the independence from $\sigma_{\rm scat}^{\rm RDP} n(t)\propto a^{-2}$ which decreases as fast as the Hubble parameter. 
So the integral depends on $T$ logarithmically. 
This means that the Universe can be transparent for the RDP without any restriction on the reheating temperature if $n \geq 2$, i.e. the RDP can be the messenger without a reheating temperature restriction. This is similar to the case of a sterile neutrino with small enough $\theta$ (see the discussion around \Eq{ycons}). The BBN and CMB constraints should be easily satisfied since they  become relevant after the electron--positron annihilation, after which the scattering probability is highly suppressed.

This messenger can be tested from observing electron recoils due to the RDP scattering in DM direct detection experiments~\cite{Zhang:2018xdp,Akerib:2019fml, Aprile:2020tmw}.  Since the experiments are sensitive to the electron recoil energy around/above $0.1\,$ keV, 
\beq \laq{eevent}\Gamma_{\rm detector}
\approx 1.2\frac{\rm events}{\rm year\cdot ton}\left(\frac{\Delta N_{\rm eff}}{0.4}\right)\left(\frac{10\,{\rm keV}}{E_{\rm{RDP}}}\right)\left(\frac{\sigma_{\rm today}}{\rm ab}\right).\eeq 
Here we assume that each target with $m_{\rm target}\approx 1\GEV$ has one electron. Detection requires less energetic RDPs, hence the RDP density can be higher. Therefore, the event rate can be enhanced compared to the nucleon case.
The keV range implies that $m_\phi/T_R \gtrsim 10^6$ can be probed. Interestingly, $\phi$ with $m_\phi \sim 10 \TEV-100\,$PeV, and decay rates to the SM as well as RDP with $\Gamma_\phi\sim \frac{m_\phi^3}{4\pi M_{\rm pl}^2}$ can be tested. This decay rate implies that $\phi$ may couple to the RDPs with Planck suppressed terms, like a modulous does.     

Let us note two things for this scenario. First, to have keV electron recoils, RDP models are severely constrained from astrophysics and cosmology (Appendix \ref{ap2}).
Second, at the considered center-of-mass energy the electron is non-relativistic, and thus, the scaling of $\s^{\rm RDP}_{\rm scat}$ is non-trivial in most realistic models. (This is also the case for nucleon-RDP scattering with $E_{\rm RDP}\lesssim 1\GEV$.)
In Appendix \ref{ap2} we sketch some models which feature suitable RDP-electron couplings.  
The event spectra by taking account of the detector efficiency and resolution will be shown in the near future.\footnote{Since the event rate is large enough, and we can evade the cosmological and astronomical constraints as discussed in Appendix.~\ref{ap2}, we may use RDPs, from modulus/inflaton decays, to explain the Xenon1T excess, similar to the boosted DM scenarios~\cite{Kannike:2020agf,Fornal:2020npv,Bloch:2020uzh}.
This may be interesting because a standard modulus decay gives a suitable keV energy scale to the RDP. The prediction of our scenario, again, is the almost isotropic RDP flux.}

\paragraph{RDP coupled to neutrinos/photons/DM/RDP}
Another option is to couple the RDP to the different types of particles making up the cosmic radiation and matter in the Universe today. 
In this case the RDP could excite the cosmic backgrounds which may provide a detectable signature. 
For instance one may consider ${\rm RDP+DM}\to \gamma \gamma$ or ${\rm RDP} \gamma\to {\rm RDP} \gamma$ reactions with $n\leq 2.$ 
While the Universe may be transparent to the RDP the rare injection of energy affects the CMB spectra due to either process. While this was discussed as a constraint for our scenario, the CMB injection may also provide for a probe of the messenger spectra.
We may also have the RDP induced cosmic-ray photons, e.g. X-rays, which could also be an interesting signature of the scenario, e.g., in DM indirect detection experiments.
Discussing this in detail as well as setting more precise BBN and CMB constraints for the messenger scatterings is left to future work.

\paragraph{Cold DM abundance}
One can even speculate that such a stabilized dark particle can also play the role of cold dark matter, given a suitable production mechanism.

One possibility is that it is produced from the thermal scattering during reheating. 
For instance, when the RDP couples to the baryon, one may expect that the dark particle is thermalized during the oscillation of 
the inflaton with $T\gtrsim 1\GEV$, when reheating is not complete. 
In that period there are sufficient amounts of baryons and anti-baryons before the annihilations.
Thus one expects the number density of the dark particle in the plasma is $n_{\rm DP}\sim T^3/\pi^2.$
The thermal relic of the number density to entropy density is diluted due to reheating as (c.f. \cite{Bezrukov:2009th,Evans:2019jcs})
$
\frac{n_{\rm DP}^{\rm cold}}{s}\sim 3\times10^{-9}\(\frac{T_R}{50\MEV}\)^{5}
$
where we have used that the plasma temperature during reheating satisfying $ g_\star \pi^2 T^4/30 \sim  \rho_\f \G_\f/H \propto a^{-3/2}$.
This mechanism provides the observed (cold) DM with $ m_{\rm RDP}\sim 60\MEV \(\frac{50\MEV}{T_R}\)^{5},$ 
obtained from $\left. (s_0/\rho_{\rm crit}) \times (m_{\rm RDP} n_{\rm DP}^{\rm cold}/s)\right|_{ T=T_R}\approx 0.12 /h^2$~\cite{Aghanim:2018eyx}, where $\rho_{\rm crit}\approx (0.03\EV)^4h^2$.
Of course, we can have other production mechanisms of the dark particle such as, e.g., the misalignment mechanism~\cite{Abbott:1982af,Preskill:1982cy,Dine:1982ah,Nelson:2011sf,Arias:2012az} when it is a scalar field.

\section{Conclusions and Discussion}

In this paper, we have studied the possibility that weakly coupled particles from inflaton decays can communicate information on the inflaton mass in relation to the reheating temperature to us, thereby acting as messengers from the reheating period.
We have shown that the neutrinos can do this job if the right-handed neutrino, satisfying the ``see-saw relation", has a mass around $\O(\MEV-\GEV)$, 
and identified the typical spectra. 
In particular, we find that such a primordial neutrino can even explain the PeV events in IceCube. While such an interpretation is speculative it shows that such particles are within reach of current experiments. 
The  hierarchy between $m_\f$ and $T_R$ can be ``naturally" predicted in a wide class of inflation models (see Appendix~\ref{ap1}).

We have also pointed out that a relativistic dark particle (RDP) from reheating, in particular a suitable right-handed neutrino, can also explain the PeV events in IceCube while alleviating the $H_0$-tension.  We can further test our scenario if more precise data on the neutrino flux, especially in the PeV region, is obtained in future observations. The information carried by more generic RDPs may also be searched for in experiments aiming at the direct or indirect detection of DM, as well as in future CMB observations.

All this nicely complements the findings of~\cite{Cicoli:2012aq,Higaki:2012ar,Conlon:2013isa,Hebecker:2014gka,Armengaud:2019uso} that axion-like particles from moduli decays are potentially observable in helioscopes such as IAXO. It shows that also other classes of particles can have suitable properties and that allow us to obtain information about the reheating phase in experiments and observations by detecting these messengers from the early Universe. 

A prediction for the messenger spectrum in general models of inflation is the almost isotropic angular dependence but with tiny anisotropies corresponding to the quantum fluctuations of the inflaton. 
Strong evidence of the reheating origin of messengers would be the correlation of the anisotropies with the ones in the CMB. But measuring this could be very challenging. 

To end let us now consider two parameter regions that we have not discussed in detail in the main text.
So far, we focused on the inflaton mass much heavier than the reheating temperature.  
Alternatively, 
the inflaton mass may be only slightly higher than the reheating temperature in certain inflaton models. 
The resulting neutrino or RDP spectra from reheating then have 
 slightly higher energy than the neutrino temperature in the current Universe. 
In particular, the ordinary spectrum of the neutrino in the energy range, $E_\nu\sim 0.1\,$eV$-1\KEV$ corresponding to $10^2\lesssim m_\f/T_R\lesssim10^7$ from \Eq{E1}, 
has a suppressed flux, and thus the background would be relatively low. As discussed around \eq{ycons}, the Universe can be transparent to $N$ for a wide range of $T_R$ depending on $\theta$. With $m_\f<10^{19}\GEV$, this region can probe a large range of $T_R$. 
An interesting possibility (other than the CMB test, see, e.g., Fig.\,\ref{fig:3}) is the PTOLEMY project~\cite{Betti:2019ouf}, in which the neutrinos from the reheating may be searched for.
The capture rate of the tritium can be estimated as~\cite{Long:2014zva,McKeen:2018xyz,Chacko:2018uke}\footnote{We have assumed that the capture cross section does not depend much on $E_\nu$ when it is much smaller than the $Q$-value of the reaction $Q\approx 18.6\KEV$. The dependence on $E_\nu$ comes from the number density of the primordial neutrino. }
\beq
 \G_{\rm detector}\sim 0.008 \frac{\rm events}{{\rm year}} \e\D N_{\rm eff} \(\frac{1\EV}{E_\nu}\) \(\frac{M_T}{ {\rm 100g}}\)
\eeq
where $M_T$ is the mass of tritium,  $\e \equiv \sum_{i=1}^3 c_i |(U_{\rm PMNS})_{e,i}|^2$, with $c_i$ being the fraction of neutrinos in the mass basis, satisfying $\sum |c_i|^2=1,$ and $U_{\rm PMNS}$ is the Pontecorvo-Maki-Nakagawa-Sakata matrix. 
Although $E_\nu \gg 1\,$eV may lead to events far from the end point of the tritium beta decay and therefore the background should be very suppressed, 
the test of reheating at PTOLEMY is, however, very challenging. Due to the constraint of $\D N_{\rm eff} \lesssim \O(0.1)$, we would need more than $\O(100)\,$kg of tritium. 

With $M_N$ and $\theta$ large enough, the decays of $N$ happen so early that the produced neutrinos scatter with the cosmic neutrino background. From the scattering neutrinos are produced at a fraction of $\O(0.1-1)$ and so a cascade happens. In this case, cascade spectra may remain until today (this has been investigated in Ref.\,\cite{Ema:2014ufa} where one can also find the typical shape of the spectra).
The shape of the cascade spectra itself may not provide a probe of reheating, since information will be lost in the multiple scatterings. 
One can nevertheless try to gain at least some information from them.
First, the spectrum depends on the time/temperature at which the neutrinos decay (cf.~\eq{cutoff}). Moreover, the higher the initial energy of the $N$ the more energy is lost from the neutrinos into other SM particles. Therefore the remaining flux is smaller, the higher the initial energy. In this way we can get information on a combination of $m_\f/T_R$ and the fraction of the total energy density of the Universe originally carried by the $N$.

\section*{Acknowledgments}
WY thanks Institut f\"ur theoretische Physik of Heidelberg University for kind hospitality when this work was initiated. WY was supported by JSPS KAKENHI Grant Number 15K21733 and 16H06490.

\appendix
\section{Inflation with a heavy inflaton and a low reheating temperature}
\label{ap1}

In presenting our main point, especially in discussing the testability of the primordial neutrino in the IceCube experiment, we took the inflaton mass to be around $M_{\rm pl}$ with $T_R\lesssim 100\GEV$. 
Let us explain which models for inflation are compatible with our purposes. However, as a general concept much wider classes of inflation models are applicable.

The original chaotic inflation model with a quadratic potential predicts an inflaton mass $m_\f\sim 10^{14}\GEV$ \cite{Linde:1983gd}. Although the original model is in tension with the limit on the tensor to scalar ratio~\cite{Akrami:2018odb}, one can slightly modify the model to get a better fit to the CMB data. 
Depending on the modification one can {have an inflaton} mass even heavier say $m_\f\sim M_{\rm pl}\approx 2.4\times 10^{18}\GEV.$ For instance one can consider models with a running kinetic term~\cite{Takahashi:2010ky},
\beq
{\cal L}= f(\f) (\partial \phi)^2-\frac{m_\f^2}{2} \f^2. \laq{HS}
\eeq
Here, we have imposed a $Z_2$ symmetry on the inflaton field $\f$ and 
$f(\f)$ is a general function of $\f$ satisfying $f(0)= 1$. 
If $f(\f)\sim {\f^{2n-2}}{n^2/M_*^{2n-2}}$ for $\f\sim M_{\rm pl}$ where inflation takes place. 
We obtain, 
\beq
{\cal L}\to  (\partial \tl{\phi})^2- \frac{m_\f^2  M_*^{2-\frac{2}{n}} }{ 2}  \tl{\phi}^{2/n} 
\eeq
by performing a field redefinition, $\tl{\f}=\f^{n}/M_*^{n-1}.$
One can then derive the condition, ({c.f. Ref.~\cite{Harigaya:2012pg}})
\beq 
\laq{masrel} (m_\f^2 M_*^{2-2/n})^{\frac{1}{(4-2/n)}}\sim10^{15-16}\GEV,
\eeq 
from the CMB normalization of the curvature perturbation. 
The spectral index, $n_s$, and the tensor to scalar ratio $r$ satisfy $n_s=1-(1+1/n)/(N_e)$ and $r =8/(N_e n)$ respectively, where $N_e=50-60$ is the number of e-folds. 
Thus, for $n> 1$, a better fit to the CMB data is obtained especially with a smaller $r$ than the quadratic term limit, $n=1$.
From \eq{masrel} one gets the inflaton mass 
\beq
\laq{massrange}
m_\f\sim 10^{13-18}\GEV  ~{\rm for }{~M_*\sim(10^{-3}-10^3)m_\f}.
\eeq
Notice that in this model the $Z_2$ symmetry stabilizes the inflaton field. 
Since the inflaton should decay to reheat the Universe, the $Z_2$ should not be an exact symmetry. 

\bigskip
An alternative possibility would be hilltop inflation~\cite{Linde:1981mu,Albrecht:1982wi} at $\f\sim 0$. In this case the potential of the inflaton may be stabilized by two or more Planck-suppressed terms such as 
\beq 
\laq{LS}
V\sim V_{\rm hilltop} -\frac{c_6\f^6}{M_{\rm pl}^2}+\frac{c_8\f^8}{M_{\rm pl}^4}+\cdots\eeq 
 at $\f \sim M_{\rm pl}.$
For instance, the potential for inflation can be given as $V_{\rm hilltop}\sim V_0-\l_\f \f^4.$ 
Although the inflation dynamics does not change due to the higher dimensional terms, 
if $c_6>0$ one cannot stabilize the inflaton field by using the dimension four and dimension six terms, both of which are usually negative and therefore decrease the potential energy by increasing $\f$ (see e.g. Refs.~\cite{Nakayama:2012dw,Guth:2018hsa, Matsui:2020wfx} for the case with $c_6<0$, where the stabilization is possible using dimension six terms). When the potential is stabilized between the higher dimensional terms, the mass 
\beq 
m_\f \sim 6.4 c_6^{2/3}c^{-1}_8 M_{\rm pl}
\eeq 
 is of the order of the Planck scale. This mass should be taken as an order of magnitude estimate, since other higher dimensional terms suppressed by $M_{\rm pl}$ also contribute at a similar order.

\paragraph{Naturalness}

Any inflaton needs to decay to SM particles to reheat the Universe. The couplings, if too large, may induce radiative corrections to the inflaton and the SM Higgs potential. Since the inflaton is heavy, the Higgs potential, especially the Higgs boson mass, would suffer from a fine-tuning requirement in order to obtain the measured one.
To avoid the tuning to the Higgs boson mass, the inflaton couplings to the SM particles should be suppressed or the correction  should  be cancelled by a symmetry such as supersymmetry. Let us discuss the former case here. In this case the reheating temperature will be much lower than the inflaton mass. 

In the Lagrangian \eq{Lag}, we can restrict $\l$, $y$, $A$, and $\l_p$ from the absence of tuning to the Higgs boson mass. In particular we want to avoid large contributions both at tree and quantum level arising from the couplings of SM particles to the inflaton and to $N$. 
From the Lagrangian, \eq{Lag}, with $\d \f =\f -\vev{\f}$, we find a bare Higgs boson mass term $\d m_h^2= -A\vev{\phi} |h|^2$.  
If this were large, we have to say that we finely tune the position of the inflaton VEV to have the observed Higgs boson mass, $m_h\approx 125\GEV$.
To avoid the tree-level tuning, we need to either identify $\d \f\approx 0 $ as an enhanced point of symmetry~[in the case of the model of \Eq{HS}], or
highly suppress $A$ satisfying $|A|\lesssim m_h^2 /|\vev{\f}|$ [in the case of the model of \Eq{LS}]. 

Let us first consider the model given in \Eq{HS} where $\f$ is stabilized at an enhanced point of symmetry, $\vev{\f}\sim 0.$
Then the breaking of $Z_2$ to have reheating may be characterized by an order parameter $\e\sim A/M_{\rm pl}$ and thus the minimum of $\f$ potential may be slightly shifted $\vev{\f}\sim \e \times M_{\rm pl}\sim \O(A). $ To avoid the tree-level tuning, we require
\beq 
m_h^2 \gtrsim \d m_h^2 \sim A^2 
\eeq 
leading to $A\lesssim \O(100\GEV).$
 Also $\l $ breaks the $Z_2$ symmetry and may be characterized by $\l \sim \O(\e).$ 
This order of magnitude estimate can be justified if $A$ is a spurion $Z_2$ odd field. Then the potential of $\f$ can have a $Z_2$ symmetric tad-pole term $\sim M_{\rm pl}^2 A \f$ which leads to $\vev{\f}= \O(A)$ with $m_\f\sim M_{
\rm pl}$. The coupling to $N$ can be from a $Z_2$ symmetric higher dimensional term, $\sim A\f \bar{N}^cN/M_{\rm pl}.$

Next let us find the conditions arising from the requirement that the radiative corrections to the Higgs boson mass from integrating out $\f$ or $N$ are smaller than $m_h$.\footnote{For the tuning argument we have not taken into account gravity loops. Devising a natural model including gravity loops or solving the hierarchy problem is way beyond the scope of our paper. }
This leads to
\begin{equation}
m_h^2\gtrsim \d m_h^2=\max{[\frac{\l_p}{16\pi^2} m_\f^2,\frac{|A|^2}{16\pi^2},\frac{|y M_N|^2}{16\pi^2},\frac{|y \l |^2}{(16\pi^2)^2}m_\f^2]}. \laq{FT}
\end{equation}
To sum up, the SM particles should be very weakly coupled to the inflaton, and therefore we get $T_R \ll m_\f$. 
This can be combined with the see-saw formula $m_\nu=(y v)^2/M_N$. 
Using an active neutrino mass $m_\nu \gtrsim \O(0.01)\EV,$ we obtain
a mass bound $M_N \lesssim 10^{7}\GEV$. 

On the other hand, for the case of the inflation model given by \Eq{LS}, there may not be an enhanced symmetry at $\vev{\f} \sim M_{\rm pl}.$ To avoid a tuning, we should take $|A|\lesssim m_h^2/|\vev{ \f}|\sim m_h^2/M_{\rm pl},$ which implies a highly suppressed $\f \to hh$ decay rate. 
To have successful reheating, one may introduce another right-handed neutrino, $\tilde{N}$, as the dominant decay product of $\f$ through $\f\to \tl{N}\tl{N}$ with a decay rate $\G_{ \f\to \tl{N}\tl{N}}.$ The SM particles are produced via the decay of $\tl{N}$. 
Since either the upper limit of the right-handed neutrino mass or the inflaton coupling to $\tl{N}$ is similarly given from the inequality of \eq{FT}, the reheating temperature is bounded from above. 
In particular if we consider the case that the $\tl{N}$ produced from the inflaton decays rapidly, $T_R \lesssim\O(100\GEV)$ for $m_\f\sim M_{\rm pl},$ due to a boost factor.
Here we have $N$ and $\tl{N}$ such that they are both mass eigenstates. 
In general we can also have decays $\f \to \tl{N} N$ whose decay rate is $\G_{\f \to N\tl{N}}$ if the inflaton couplings are not aligned in the mass eigenbasis. This is equivalent to taking account of $N-\tl{N}$ oscillations.  
The amount of produced $N$ is determined from both $\G_{\f\to N\tl{N}}/2$ and $\G_{\f \to NN}.$
If decays $\tl{N}$ promptly, all of discussions in the main part remain unchanged by taking the replacement of $\G_{\f\to hh}$ with $\G_{\f \to \tl{N}\tl{N}}$ and $\G_{\f \to NN}$ with $\G_{\f\to N\tl{N}}/2+\G_{ \f \to NN}$. 

\section{Constraints on RDPs and models of keV electron recoils caused by RDPs}
\label{ap2}
Here, we discuss cosmological and astrophysical constraints on an RDP-electron coupling with $E_{\rm RDP}\sim \KEV$. We also 
 briefly discuss consistent models for the constraints, and a UV model with a concrete electron-RDP cross section. The cross section has an energy dependence such that the Universe is transparent to the RDP independent of the reheating temperature. 
For clarity we denote the dark particle produced from other processes than inflaton decays as DP. 

\paragraph{Constraints at keV and DP mass}
Let us consider processes with an energy scale around keVs, where the relevant cross sections can be expected to be around $\s_{\rm today}.$ 
The cores of red giant, white dwarf and horizontal branch stars have temperatures around $\O(1-10)$keV. 
New weakly coupled particles could induce an extra cooling in these objects by emitting weakly coupled DPs. 
A relevant process for this would be a Compton scattering- or Bremsstrahlung-like effect, $e \gamma\to e +{\rm 2 DP}$~\cite{Raffelt:1996wa,Dreiner:2013tja}.
We can obtain a simplistic estimate by comparing to the limits obtained for an ALP/axion via $e \gamma \to e a$. In this situation there is a bound on the axion electron coupling~\cite{Viaux:2013lha,Bertolami:2014wua,Capozzi:2020cbu} $g_{aee}<\O(10^{-13})$.
Let us write an effective coupling, 
\begin{equation}
    g_{aee}^{\rm eff} \sim \sqrt{\sigma_{\rm today}} m_e
\end{equation} 
for the vertex of $e\to e+$2DP, i.e. we assume a factorization of the amplitude~\cite{Raffelt:1996wa}, which should hold since the photon momentum is soft compared with the electron mass. 
From 
$g^{\rm eff}_{aee}<\O(10^{-13})$, we then obtain 
\beq 
\sigma_{\rm today}\lesssim \, \O(10^{-4}){\rm ab} ~~{\rm if ~m_{\rm DP}\lesssim 10\KEV}.
\eeq 
For larger $m_{\rm DP}$ this bound is not applicable because efficient production requires the mass to be smaller than the typical temperature of the cores. 
In principle, there are also other constraints, e.g. from loop-induced photon couplings generating an emission from horizontal branch, red-giant, white dwarf stars or the sun~\cite{Raffelt:1988rx,Raffelt:1996wa}. However, these would require a more careful analysis and concrete model realizations. 

One simple possibility to be consistent with this constraint is to set the mass of the DP above the typical temperatures, $m_{\rm DP}\gtrsim \O(10)\KEV$. 
Therefore we focus on\footnote{If one can precisely measure the momentum of a DP, from a non-relativistic DP one could, in principle, also probe the reheating.} 
\beq 
E_{\rm RDP}\gtrsim m_{\rm DP}\gtrsim \O(10)\KEV.
\eeq
The resulting recoil energy of electron should be $\gtrsim \O(0.1)\KEV$.

\paragraph{Constraints at MeV energies and a consistent model of a messenger}
In the early Universe, electrons and positrons can annihilate to the DP pairs. Let us denote the annihilation cross section at $T\sim m_e$ as $\sigma_{\rm MeV}$. 
Slightly above this temperature the annihilation rate is  
$\G_{\rm MeV}\sim \s_{\rm MeV} T^3/\pi^2 $. This should be smaller than the Hubble expansion rate to have consistent BBN~\cite{Sabti:2019mhn, Chigusa:2020bgq}. This requires 
\beq
\laq{BBN}
\s_{\rm MeV}\lesssim  \sigma_{\rm BBN}=\O(1-10){\rm ab}.
\eeq

As we can see at MeV energies the bound is much stricter than the one obtained for the cross section today. To obtain a suitable scaling of the cross section with the energy, we may introduce a light mediator, $\tl{\f}$, whose mass satisfies
\begin{equation}
    \O(10)\KEV\lesssim m_{\tl{\f}}
    \lesssim m_e,
\end{equation}
where we take the mediator mass to be large enough to alleviate the stellar cooling constraints. 

A suitable interaction is 
\beq 
{\cal L}_{\rm UV}=-\tilde{Y} \tilde{\phi} \bar{e}e- \tilde{A}\tilde{\f} \F_{\rm DP}^2.
\eeq 
Here, $\F_{\rm DP}$ is the DP which is a real scalar. $\tl{Y}$ is a Yukawa coupling, and $\tl{A}$ is a dimension one coupling. 
With this the cross section of DP--electron scattering can be obtained as 
\begin{align}
\s_{\rm scat}^{\rm RDP}&\propto  \frac{\tl{Y}^2 \tl{A}^2}{4\pi E_{\rm RDP}^4} ~~{{\rm if}  ~~m_{\tl{\f}} \lesssim E_{\rm RDP}\lesssim m_e}\\ &\propto \frac{\tl{Y}^2 \tl{A}^2}{4\pi E_{\rm cm}^4} ~~~~{{\rm if}  ~~m_e \lesssim E_{\rm RDP}}.
\end{align}
The cross section scales as $E_{\rm RDP}^{-4}$ with $E_{\rm RDP}<m_e$ because of the t-channel momentum exchange of $\O(E_{\rm RDP})$.
At higher center-of-mass energy $E_{\rm cm}\sim m_e E_{\rm RDP}$, it scales as $E_{\rm RDP}^{-2}$. This means $n=4$ in \Eq{scatDM}. From~\eq{lepton}, we find that the Universe can then easily be sufficiently transparent to it.

Setting, for simplicity, $m_{\tl{\f}}\sim m_{\rm DP}$,  we get (by assuming the scattering and annihilation cross sections are of the same order at a MeV center-of-mass energy)
$
\sigma_{\rm today}\sim \sigma_{\rm MeV} (\frac{m_e}{E_{\rm RDP}})^4 \lesssim \O(10-100) {\rm nb}  \(\frac{10\KEV}{E_{\rm RDP}}\)^2.
$

In this concrete model a more severe bound arises from BBN, in particular from the process $e+\bar{e}\to \tl{\f}+\gamma$. This has a cross section of the order of \begin{equation}
   \s_{e\bar{e}\to \tl{\f}\g}|_{T\sim m_e}\sim \frac{\tl{Y}^2 e^2}{m_e^2}\sim  \s_{\rm today} {e^2}{(E_{\rm RDP}/\tl{A})^2} (E_{\rm RDP}/m_e)^2 .
\end{equation}
Noting that $\tl{A}\lesssim m_{\f}\sim m_{\tl{\f}}\lesssim E_{\rm RDP}$ is needed to have a stable vacuum for $\tl{\f} -\F_{\rm DP}$ system and requiring $\s_{e\bar{e}\to \tl{\f}\g}|_{T\sim m_e}\lesssim \sigma_{\rm BBN}$ we find
\beq
\laq{BBNcons}
\s_{\rm today}\lesssim \O(0.1-1){\rm pb} \times \(\frac{10 \KEV}{E_{\rm ERP}}\)^2 ~[{\rm BBN ~bound}].
\eeq
The inequality saturates when $E_{\rm RDP} \sim m_\phi \sim m_{\tl{\phi}}\sim \tl{A}$ and 
\beq 
\tl{Y}\sim 10^{-10}.
\eeq

In the following we consider a small $\tl{Y}$ but not too small  $\tl{A}/m_{\tl{\phi}}.$
In this case, 
the constraints from ground-based experiments, such as beam dump experiments (cf., e.g.~\cite{Beacham:2019nyx}), can be satisfied. 
Even if $\tl{\phi}$ is produced in an experiment, it dominantly decays to DPs without being detected. 

We may also worry that such a system is constrained from supernovae~\cite{Raffelt:1996wa,Dreiner:2013mua,DeRocco:2019jti}.
For higher energy scales, the relevant cross section to $ee^+ \to 2\F_{\rm DP}\OR e\F_{\rm DP}\to e\F_{\rm DP} $ is highly suppressed by $1/E_{\rm cm}^4$ in the center-of-mass frame. 
This alleviates the constraints for the relevant energy-loss process in supernovae. 
However, we again note that, $e \tl{\f}\to e \gamma, e\bar{e}\to \tl{\f}\gamma $ processes, the cross section of which is $\s_{e\bar{e}\to \tl{\f}\g}\sim e^2 (E_{\rm RDP}/\tl{A})^{2} (E_{\rm RDP}/E_{\rm cm})^2 \sigma_{\rm today},$ may also 
contribute to the energy-loss.
Although this process is usually not considered for axions, it may be important for this model. 
The so-called Raffelt criterion sets the upper bound on the cross-section to emit a weakly coupled particle as
~\cite{Raffelt:1996wa, Dreiner:2013mua}
\beq
\(\frac{T_c^3}{\pi^2}\)\s_{e\bar{e}\to \tl{\f}\g}|_{T\sim T_c} \lesssim 10^{19}{\rm erg\cdot g^{-1}\cdot s^{-1}}
\eeq
where the SN1987 core temperature is considered as $T\equiv T_c \sim 35\MEV$. 
This leads to a severe bound \beq\s_{e\bar{e}\to \tl{\f}\g}|_{T\sim T_c}  \lesssim \s^{\rm cooling}_{\rm SN}\equiv 1{\rm yb}.\eeq
However, taking the scaling into account, at low energy scales, it is not quite so severe, 
\beq
\label{eq:supernovae}
\sigma_{\rm today}\lesssim 0.1 {\rm fb} \(\frac{10\KEV}{E_{\rm RDP}}\)^{2} ~~[{\rm SN1987}]
\eeq 
This still allows $\O(100)$ events per year from~\eq{eevent}, although we hasten to point out that our estimates are very rough.\footnote{We could also couple $\tl{\f}$ to nucleons (and $\F_{\rm DP}$) so strongly that $\F_{\rm DP} \AND \tl{\f}$ are trapped in the core of the supernova or a neutron star. 
In this case the constraint from energy loss in supernovae may be evaded. }

To summarize, we found in a model with two additional scalars, the dark particle $\F_{\rm DP}$ and the mediator $\tl{\f}$ heavier than $10\KEV$, the electron-RDP scattering cross section today is dominantly constrained from BBN, \eq{BBNcons}, 
and supernovae physics, Eq.~\eqref{eq:supernovae}. 
Therefore for a few tons of target material, we can have much more than 1 event per year. 
We note that our conclusion differs from that of Ref.~\cite{Chigusa:2020bgq}, because of the different energy dependence of the cross-sections. 
The light mediator, $\tl{\phi}$ in our model, makes various cross sections smaller, and thus constraints weaker at higher energy scales. 
That said, we emphasize that we only provide very simplistic estimates and a more detailed analysis could constrain the model more strongly.

\bibliographystyle{utphys}
\bibliography{references}

\end{document}